\documentclass[11pt]{article}

\usepackage[preprint]{acl}

\usepackage{times}
\usepackage{latexsym}
\usepackage[T1]{fontenc}
\usepackage[utf8]{inputenc}
\usepackage{amsmath}
\usepackage{amssymb}
\usepackage{amsthm}
\newtheorem{definition}{Definition}
\usepackage{microtype}
\usepackage{inconsolata}
\usepackage{graphicx}
\usepackage{subcaption}
\usepackage{booktabs}
\usepackage{multirow}
\usepackage{colortbl}
\usepackage{xcolor}
\usepackage{tabularx}
\usepackage{enumitem}
\usepackage{placeins}
\usepackage{algorithm}
\usepackage{algpseudocode}

\title{From Saliency to Discriminability: Rank-Preserving Visual Token Pruning for VLM Rerankers}

\author{
  \textbf{Siyi Liu}\textsuperscript{1,$\dagger$} \quad
  \textbf{Hanjun Yang}\textsuperscript{1,$\dagger$} \quad
  \textbf{Chenchen Zhang}\textsuperscript{1} \quad
  \textbf{Xiaorong Zhu}\textsuperscript{1} \quad
  \textbf{Xinyu Zuo}\textsuperscript{2} \\
  \textbf{Lisheng Duan}\textsuperscript{2} \quad
  \textbf{Haijin Liang}\textsuperscript{2} \quad
  \textbf{Jin Ma}\textsuperscript{2} \quad
  \textbf{Junfu Pu}\textsuperscript{3} \quad
  \textbf{Yongqi Zhang}\textsuperscript{1,*} \\
  \normalfont
  \textsuperscript{1}The Hong Kong University of Science and Technology (Guangzhou) \\
  \textsuperscript{2}Tencent Yuanbao \quad
  \textsuperscript{3}ARC Lab, Tencent \\
  \texttt{ssui.liu1022@gmail.com \quad hyang371@connect.hkust-gz.edu.cn} \\
  \texttt{yongqizhang@hkust-gz.edu.cn}
}

\begin{document}
\maketitle
\begingroup
\renewcommand{\thefootnote}{$\dagger$}
\footnotetext{Equal contribution.}
\renewcommand{\thefootnote}{*}
\footnotetext{Corresponding author.}
\endgroup

\begin{abstract}
Large vision-language models used as listwise rerankers must jointly process visual tokens from tens of candidates per query, making token pruning essential for practical deployment. Existing pruning methods retain tokens by attention saliency, yet we show that saliency is systematically misaligned with ranking contribution: visually prominent tokens often capture order-neutral patterns shared across candidates. This mismatch is layer-dependent: saliency becomes informative only where attention is concentrated, and normalized attention entropy diagnoses the reliability shift (Pearson $r{=}0.87$). We propose \textbf{RaDiCal} (\textbf{Ra}nk-\textbf{Di}scriminative \textbf{Cal}ibration), a training-free framework that uses normalized attention entropy to decide when saliency can be trusted, fusing it with an attention-free rank-discriminative prior and selecting pruning layers from the same trust landscape. Across three retrieval benchmarks and multiple VLM architectures, RaDiCal matches Dense MRR@10 on Flickr30K and surpasses it on MSCOCO at a 20\% token budget, ranks first among all pruning methods on FashionIQ, and holds within 1.2\,pp on Flickr30K and MSCOCO at 10\% retention. It cuts FLOPs by 39--45\% and delivers 1.28--1.45$\times$ measured speedups across two VLM architectures without dataset-specific retuning.
\end{abstract}

\section{Introduction}

\begin{figure}[!t]
\centering
\includegraphics[width=\linewidth]{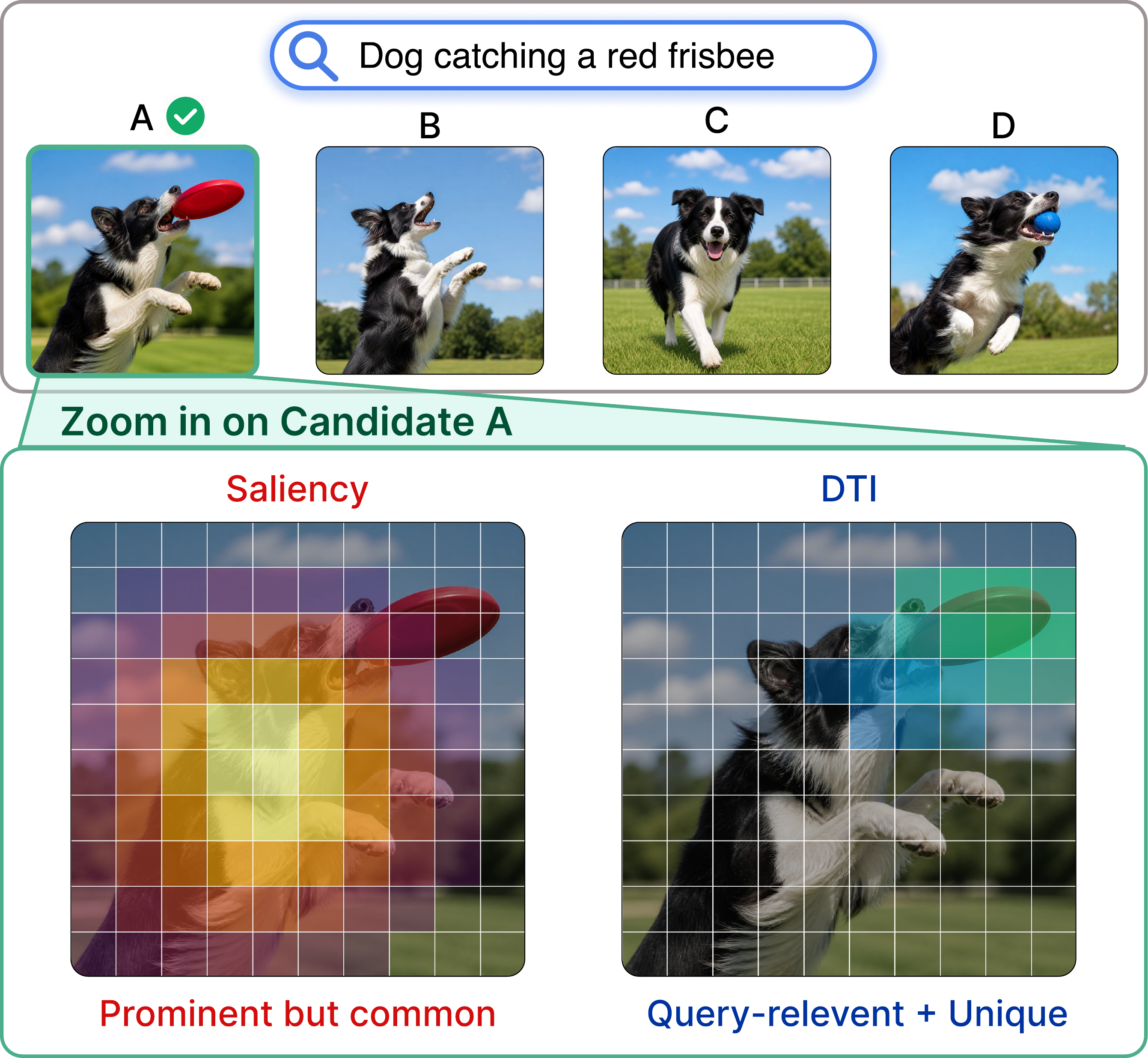}
\caption{Saliency highlights visually prominent but common regions, while rank-discriminative scoring identifies query-relevant, candidate-distinctive evidence for reranking.}
\label{fig:teaser}
\end{figure}

Following LLM-based listwise reranking~\cite{rankgpt,icr}, large vision-language models increasingly serve as listwise rerankers~\cite{lamra,cotrr,unimev2,ranknexus}, jointly scoring a query against tens of candidate images in one forward pass. This setting makes visual tokens a dominant inference bottleneck: each high-resolution image may contribute hundreds or thousands of visual tokens \cite{inferenceoptimalvlms,sparsevlm,visionzip}, and listwise reranking multiplies this cost by tens of candidates. Token pruning is therefore essential for practical VLM reranking, but compression here demands a different fidelity criterion: a reranker produces an ordered list rather than a complete image description, so the key question is whether pruning preserves query-conditioned candidate order.

Most visual token pruning methods follow a saliency-preservation view: they retain tokens appearing important within an image or a single multimodal input \cite{fastv,llavaprumerge,pyramiddrop,sparsevlm,visionzip,pvc,divprune}. This criterion is natural for single-image understanding but misaligned with listwise reranking. Consider the query ``a dog catching a red frisbee'' (Figure~\ref{fig:teaser}): a saliency-based pruner may focus on the dog body, visually prominent and repeatedly attended to, yet if most candidates contain a dog, this region provides little evidence for ordering them~\citep{cdpruner}. The rank-relevant evidence is instead the red frisbee and the catching interaction, both query-relevant and discriminative among visually similar candidates.

In our diagnostic study (\S\ref{sec:preliminary}), attention saliency is unreliable for ranking at multiple layers: saliency--ranking alignment is indistinguishable from a random-ordering null. Yet the failure is not uniform across layers \cite{tokenpruning_worse_random,devilsmiddlelayers,pyramiddrop}: saliency becomes informative where attention is concentrated but remains unreliable where it is diffuse, calling for calibration rather than wholesale discard.

A token is rank-critical only when it is relevant to the query and distinguishes its candidate from others. However, a ranking-specific prior alone is insufficient: saliency can still provide useful evidence when concentrated but is unreliable when diffuse. Rank-preserving pruning therefore requires three coupled decisions: which visual evidence should be preserved (\textbf{what}), how much attention saliency should be trusted at each layer (\textbf{when}), and which layers should perform pruning under this trust landscape (\textbf{where}).

We propose \textbf{RaDiCal} (\textbf{Rank-Discriminative Calibration}), a training-free visual token pruning framework for VLM rerankers. RaDiCal computes \textbf{Discriminative Token Importance (DTI)}, an attention-free prior combining query relevance with cross-candidate distinctiveness. It fuses DTI with \textbf{AttentionInfo}, a layer-specific information-theoretic saliency, through a trust coefficient derived from normalized attention entropy~\cite{agilepruner,coast}. \textbf{$\alpha$-Maximin} selects pruning layers from the same entropy curve, spanning diverse trust regimes without dataset-specific sweeps.

Our contributions are as follows:
\begin{itemize}[leftmargin=*]
    \item \textbf{Ranking-specific diagnosis.}
    We identify a systematic mismatch between attention saliency and ranking contribution in VLM-based listwise reranking. This mismatch is layer-dependent: normalized attention entropy diagnoses where saliency becomes trustworthy, motivating calibrated rather than unconditional use of attention.

    \item \textbf{Rank-discriminative token value.}
    We introduce \textbf{Discriminative Token Importance (DTI)}, an attention-free prior that scores each token by query relevance and cross-candidate distinctiveness, shifting pruning from saliency preservation to rank-critical evidence preservation.

    \item \textbf{Trust-calibrated pruning.}
    We propose \textbf{RaDiCal}, a training-free framework that calibrates saliency with normalized attention entropy and uses the resulting reliability signal for both token-score fusion and pruning-layer selection, eliminating dataset-specific sweeps.

    \item \textbf{Empirical validation.}
    Across three benchmarks and two VLM architectures, RaDiCal matches or exceeds Dense on Flickr30K and MSCOCO at a 20\% token budget, ranks first among all pruning methods on FashionIQ, delivers measured wall-clock speedups of up to 1.45$\times$, and transfers across datasets, model scales, and architectures without retuning.
\end{itemize}

\section{Related Work}
\label{sec:related_work}

\subsection{VLM Reranking}
\label{sec:rw_reranking}

VLMs serve as reranking modules in retrieval pipelines \cite{lamra,mmembed,unimev2}, operating in pointwise \cite{unimev2,ragvl} or listwise mode \cite{lamra,cotrr,ranknexus}; listwise methods evaluate candidates relationally within a shared context. Each image contributes hundreds to thousands of visual tokens, and visual-token pruning can substantially reduce this cost on single-image benchmarks \cite{fastv,sparsevlm,pyramiddrop,visionzip}, with recent reranking work also exploiting input compression \cite{ziprerank}---but token-value criteria remain designed for single-image fidelity, not candidate ordering.

\subsection{Visual Token Pruning}
\label{sec:rw_saliency}

Attention-based pruning is effective for single-image tasks \cite{fastv,sparsevlm,pyramiddrop}, and layer-adaptive or complexity-adaptive schedules further improve it \cite{pyramiddrop,atpllava,adaptinfer,hidrop}. Yet attention does not equal contribution: low-attention tokens affect outputs \cite{beyondintermediatestates}, high-attention tokens serve as probability dumps \cite{capa}, and random pruning can match designed methods \cite{vtcbench}. These results reframe saliency as an empirical hypothesis whose validity depends on context.

Saliency reliability varies sharply across layers: an information horizon shifts with task complexity \cite{tokenpruning_worse_random}, middle layers show abrupt reliability changes \cite{devilsmiddlelayers}, and positional biases distort multi-image attention \cite{feather}. Attention entropy indicates when saliency is informative \cite{agilepruner,coast}, though it remains a weak correctness predictor in isolation \cite{wherereliabilitylives} and has been applied only to single-image tasks; in text-only LLM QA, supervised head-entropy features do predict answer correctness~\cite{ostmeier_entropy}, but neither setting addresses unsupervised saliency calibration across competing images. In listwise reranking, the challenge compounds: token importance must be judged across both layers and competing candidates, where shared visual patterns inflate saliency for rank-irrelevant tokens.

\subsection{Ranking-Aware Compression}
\label{sec:rw_gap}

Training-free pruning can transfer across datasets via attention-derived schedules \cite{fitprune}, but the token-value criterion remains single-input preservation. Multi-image approaches exploit inter-image redundancy or diversity \cite{pvc,tofu,trimtokenatorlc}, with recent work balancing importance and diversity \cite{idpruner}; prompt-level layouts enable comparison above the token-selection layer \cite{square}; video methods leverage temporal continuity \cite{dycoke,surge}---absent in listwise reranking with discrete, unordered candidates. Ranking-aware passage compression exists for text retrieval \cite{c2r,perank,oscar}, but visual saliency introduces the conditional-reliability issues above.

\subsection{Broader Efficiency Paradigms}
\label{sec:rw_broader}

Beyond the attention-saliency pruning~\cite{fastv,pyramiddrop} and redundancy-aware merging~\cite{sparsevlm,visionzip,dart} methods discussed above, other work cuts visual cost upstream of token selection:
CARES~\cite{cares} and AwaRes~\cite{awares} adapt input resolution per query via a trained classifier or tool-calling with RL;
CROP~\cite{crop} and ERGO~\cite{ergo} train localization modules for region-oriented selective pruning;
and VisionThink~\cite{visionthink} trains an RL policy that decides at inference time whether to request the full-resolution image.
All of these methods operate on single-image inputs with task-specific training or learned policies; none targets cross-candidate rank preservation under a shared listwise context.

\section{Problem Formulation and Diagnostics}
\label{sec:preliminary}

\begin{figure*}[!t]
\centering
\begin{subfigure}[b]{0.64\textwidth}
    \centering
    \includegraphics[width=\textwidth]{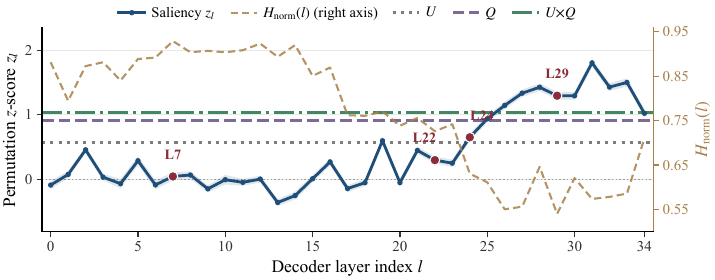}
    \caption{Layer-wise saliency reliability curve.}
    \label{fig:preliminary_curve}
\end{subfigure}
\hfill
\begin{subfigure}[b]{0.34\textwidth}
    \centering
    \includegraphics[width=\textwidth]{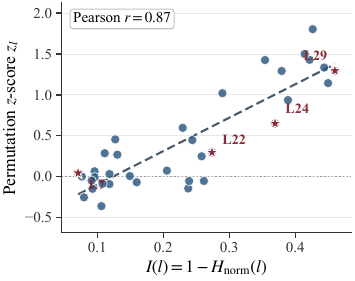}
    \caption{Information density scatter.}
    \label{fig:preliminary_scatter}
\end{subfigure}
\caption{Preliminary diagnostic of layer-conditioned saliency reliability.
(a)~$z_l$ is layer-dependent; $H_{\text{norm}}$ overlaid on right axis, attention-free baselines (U, Q, U$\times$Q) as dashed lines.
(b)~Reliability correlates with information density $I(l){=}1{-}H_{\text{norm}}(l)$ (Pearson $r{=}0.87$).
See \S\ref{sec:saliency_failure} for definitions.}
\label{fig:preliminary}
\end{figure*}

\subsection{Saliency Reliability Is Layer-Conditioned}
\label{sec:saliency_failure}

\paragraph{Setup.} Consider a text query $q$ and a candidate set $\mathcal{C} = \{c_1, \dots, c_N\}$, where candidate $c_k$ is encoded by a vision encoder into $M_k$ visual tokens, for a total of $V = \sum_{k=1}^{N} M_k$ visual tokens. A VLM processes the query and candidate visual tokens in a shared multimodal context to produce ranking scores. The pruning objective is \emph{ranking preservation}: removing visual tokens while maintaining relative candidate order. We measure each layer's saliency reliability via a permutation-corrected $z$-score $z_l$: the extent to which saliency--ranking alignment exceeds a random-ordering null (protocol in Appendix~\ref{app:diagnostic_protocol}).

\paragraph{Per-layer reliability curve.} Figure~\ref{fig:preliminary_curve} plots $z_l$ across all decoder layers. At shallow layers (e.g., L7), $z_l \approx 0$, but this global failure masks finer layer-dependent structure. The reliability profile is strongly non-monotonic: specific middle and deep layers (e.g., L22, L24, L29) recover substantially stronger alignment. This non-monotonicity reflects the listwise nature of the task: shared foreground patterns can attract attention without affecting relative order, while query-specific differences become useful only after text--visual alignment sharpens. Dashed lines mark attention-free reference baselines (U, Q, U$\times$Q) from vision-encoder embeddings; U$\times$Q exceeds saliency at several high-$H_{\text{norm}}$ layers, showing that embedding-space signals complement saliency where attention is uninformative (formalized in Definition~\ref{def:rank_discriminative}).

A fixed pruning layer --- the default in prior work~\citep{fastv,pyramiddrop} --- will either prune where saliency is unreliable or miss the window where it becomes informative~\citep{adaptinfer}.

\subsection{Entropy as a Trust Signal}
\label{sec:hnorm}

\paragraph{Normalized attention entropy.} We define $H_{\text{norm}}$ to quantify how discriminative each layer's attention is. Let $A^{(l)}$ denote the full attention weights at layer $l$. We first restrict each text row to the $V$ visual-token columns and renormalize within those columns:
\begin{equation}
\begin{aligned}
\bar{A}^{(l)}_{r,i} &=
\frac{A^{(l)}_{r,i}}{\sum_{j=1}^{V} A^{(l)}_{r,j}}, \\
p^{(l)}_i &= \frac{1}{N_t}\sum\nolimits_{r=1}^{N_t} \bar{A}^{(l)}_{r,i},
\quad \sum\nolimits_{i=1}^V p^{(l)}_i = 1.
\end{aligned}
\end{equation}
Its normalized Shannon entropy is:
\begin{equation}
H_{\text{norm}}(l) = \frac{-\sum_{i=1}^V p^{(l)}_i \log p^{(l)}_i}{\log V}.
\end{equation}
$H_{\text{norm}} \approx 1$ indicates near-uniform attention with little discriminative signal; $H_{\text{norm}} \approx 0$ indicates concentrated attention where saliency is informative~\citep{agilepruner}. Equivalently, $1 - H_{\text{norm}}(l)$ measures the normalized information gain over a uniform prior; we formalize this connection in \S\ref{sec:method}.

\paragraph{Layer-wise entropy profile.} Across decoder layers, $H_{\text{norm}}$ is non-monotonic---transitioning from a high-entropy shallow plateau through a descent ramp to a deep valley near L29, with a partial rebound in the final layers---and model-intrinsic: Flickr30K and MSCOCO yield Pearson $r = 0.9993$, requiring no per-dataset recalibration.

\paragraph{Reliability correlates with information density.} Figure~\ref{fig:preliminary_scatter} plots $z_l$ against information density $I(l) = 1 - H_{\text{norm}}(l)$. The strong positive relationship (Pearson $r = 0.87$ across 36 decoder layers) shows that layer-wise reliability is governed by attention concentration: more focused attention yields saliency better aligned with ranking contribution. This turns the layer-conditioned variation observed in Panel~(a) into a reliable calibration signal: $H_{\text{norm}}$ serves as a lightweight, inference-time proxy for how much to trust saliency at each layer.

\paragraph{Design implications.}
\label{sec:design_questions}
These diagnostics motivate three requirements for rank-preserving pruning: an attention-free token prior for layers where saliency is unreliable, a calibration mechanism that adapts fusion to layer-wise trust, and a layer schedule spanning diverse reliability regimes---all addressed in \S\ref{sec:method}.

\section{Method: RaDiCal}
\label{sec:method}

\subsection{Overview}
\label{sec:method_overview}

\textbf{RaDiCal} (\textbf{Ra}nk-\textbf{Di}scriminative \textbf{Cal}ibration) is a pruning framework in which $H_{\text{norm}}$ simultaneously governs signal fusion and layer selection through a single coordinate (Figure~\ref{fig:pipeline}). Two complementary token scores---a layer-invariant DTI prior (\S\ref{sec:dti}) and a layer-specific AttentionInfo saliency (\S\ref{sec:attinfo})---are geometrically fused via a trust coefficient $\alpha(l)$ derived from $H_{\text{norm}}$ (\S\ref{sec:trust_fusion}), which also drives $\alpha$-Maximin layer scheduling (\S\ref{sec:scheduling}).

\subsection{Complementary Token-Level Signals}
\label{sec:token_signals}

We construct two complementary token-level signals: DTI captures query-conditioned cross-candidate discriminativeness (layer-invariant), and AttentionInfo captures layer-specific information content (layer-dependent).

\subsubsection{Discriminative Token Importance}
\label{sec:dti}

Let each candidate $c_k$ be encoded by the vision encoder into visual tokens $\{t_i^k\}_{i=1}^{M_k}$, where $M_k$ may vary across candidates and $V = \sum_{k=1}^{N} M_k$ denotes the total number of visual tokens.

\begin{definition}[Rank-discriminative score]
\label{def:rank_discriminative}
Given a query $q$, a candidate set $\mathcal{C}=\{c_1,\ldots,c_N\}$, and visual tokens $\{t_i^k\}_{i=1}^{M_k}$ for each candidate $c_k$, the rank-discriminative score of token $t_i^k$ captures both cross-candidate distinctiveness and query relevance. We instantiate these as:
\begin{align*}
\mathrm{CCU}(t_i^k,\mathcal{C}) &= 1-\cos(t_i^k,\,\bar{v}), \\
\mathrm{QRel}(t_i^k,q) &= \max(\cos(t_i^k,\,q_{\mathrm{emb}}),\;0),
\end{align*}
where $\bar{v}$ is the L2-normalized mean of all candidate visual tokens (global prototype) and $q_{\mathrm{emb}}$ is the query text embedding. The Discriminative Token Importance is then:
\begin{equation}
\mathrm{DTI}(t_i^k) \;\propto\; \mathrm{CCU}(t_i^k,\mathcal{C}) \cdot \mathrm{QRel}(t_i^k,q).
\end{equation}
\end{definition}

This definition separates ranking-relevant token value from within-image saliency: a token must be both query-relevant \emph{and} cross-candidate distinctive to affect candidate ordering. DTI is computed once from ViT outputs before the LLM forward pass and reused across all selected layers; its runtime cost is discussed in \S\ref{sec:efficiency} and Appendix~\ref{app:efficiency}.

\begin{figure}[!t]
    \centering
    \includegraphics[width=\columnwidth]{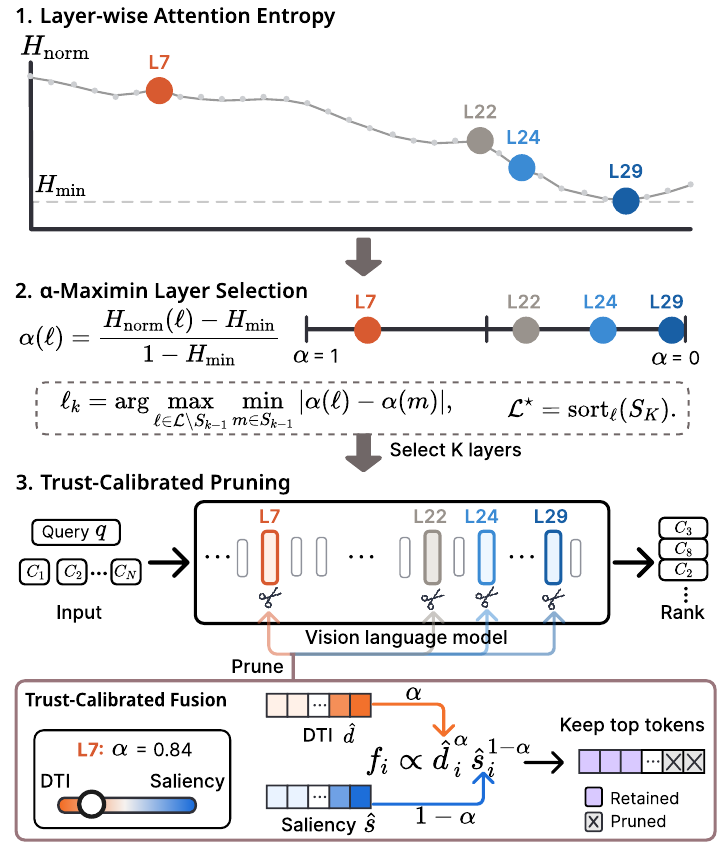}
    \caption{RaDiCal pipeline. (1)~$H_{\text{norm}}$ profile across layers identifies attention concentration patterns. (2)~$\alpha$-Maximin maps $H_{\text{norm}}$ to trust $\alpha(l)$ and selects $K$ pruning layers spanning diverse trust regimes. (3)~At each selected layer, DTI and saliency are geometrically fused via $\alpha$ to retain top tokens.}
    \label{fig:pipeline}
\end{figure}

\subsubsection{AttentionInfo Saliency}
\label{sec:attinfo}

DTI is layer-invariant and does not capture how attention concentration varies across the LLM pipeline. From the information-gain view of $H_{\text{norm}}$ (\S\ref{sec:hnorm}), AttentionInfo decomposes the layer-level KL gap to individual tokens (derivation in Appendix~\ref{app:attinfo_derivation}). Let $T$ denote the current active token set with $V_T = |T|$ tokens, and $p^{(l,T)}$ the text-averaged attention distribution over active visual tokens at layer $l$. The token-level saliency is:
\begin{equation}
\text{AttInfo}^{(l,T)}_i = \max\!\bigl(p^{(l,T)}_i \cdot \log(V_T \cdot p^{(l,T)}_i),\; 0\bigr).
\end{equation}
Token $i$ receives a positive score if and only if its attention weight exceeds the active-set uniform baseline $1/V_T$. Normalizing along the active token dimension yields $\text{Sal}^{(l,T)}_i \in [0,1]$, the saliency input to \S\ref{sec:trust_fusion}.

\subsection{Trust-Calibrated Fusion}
\label{sec:trust_fusion}

The trust trade-off between DTI and AttentionInfo varies with $H_{\text{norm}}$: high entropy favors DTI, low entropy favors saliency.

\paragraph{$\alpha$: from $H_{\text{norm}}$ to a per-layer trust weight.}
\label{sec:alpha}

Let $H_{\min}$ denote the model's minimum $H_{\text{norm}}$ value, the normalized entropy at the layer where attention is most concentrated (established in \S\ref{sec:hnorm}). We define the per-layer trust coefficient:
\begin{equation}
\alpha(l) = \frac{H_{\text{norm}}(l) - H_{\min}}{1 - H_{\min}} \;\in\; [0,\,1].
\end{equation}
At $\alpha(l) = 0$ ($H_{\text{norm}}(l) = H_{\min}$, sharpest layer), saliency is most reliable and AttentionInfo fully governs token selection. At $\alpha(l) = 1$ ($H_{\text{norm}}(l) = 1$, uniform attention), saliency is uninformative and DTI, the attention-free prior, governs entirely. $H_{\min}$ is read from the model's own $H_{\text{norm}}$ curve (computed offline; see \S\ref{sec:pipeline} and Figure~\ref{fig:pipeline}, Step~1); no dataset-specific calibration is required.

\subsubsection{Geometric Fusion}
\label{sec:geometric_fusion}

Given $\alpha(l)$, normalized DTI scores $\hat{d}_i \in [0,1]$, and normalized AttentionInfo scores $\hat{s}^{(l)}_i \in [0,1]$ (\S\ref{sec:attinfo}), we compute:
\begin{equation}
f_i^{(l)} \;\propto\; \max(\hat{d}_i,\epsilon)^{\alpha(l)} \;\cdot\; \max(\hat{s}_i^{(l)},\epsilon)^{1-\alpha(l)}.
\end{equation}
This log-linear form provides \emph{scale-invariance} (DTI and Sal need not share a numerical scale) and, for $\alpha(l) \in (0,1)$, \emph{soft-veto} behavior: a near-zero value in either channel suppresses $f_i^{(l)}$. At the endpoints the fusion degenerates to the surviving channel by design ($\alpha{=}0$: pure saliency; $\alpha{=}1$: pure DTI).

\subsection{Layer Scheduling}
\label{sec:scheduling}

The same $\alpha$ curve also selects which layers should prune, extending trust calibration from token scoring to layer scheduling. $\alpha$-Maximin (\S\ref{sec:maximin}) operates on the $\alpha$ curve defined in \S\ref{sec:alpha}. Given a chosen schedule size $K$, a fixed layer-gap guard $g$, and a global keep ratio $R$, the scheduler produces an ordered layer set $\mathcal{L}^*$.

\paragraph{$\alpha$-Maximin layer selection.}
\label{sec:maximin}

A schedule of $K$ pruning layers should span diverse trust regimes. Clustering selected layers near the same $\alpha$ value wastes the available trust spectrum, as adjacent-$\alpha$ layers yield redundant pruning decisions~\citep{pyramiddrop}. We therefore use greedy farthest-first selection in $\alpha$-space, anchored at the most reliable saliency layer, with a minimum layer-gap guard $g$ (Figure~\ref{fig:pipeline}, Step~2; tie-breaking and relaxation details in Algorithm~\ref{alg:trust_pruning}).

\begin{table*}[!t]
\centering
\small
\setlength{\tabcolsep}{2.5pt}
\begin{tabular}{@{}l cccc cccc cccc@{}}
\toprule
& \multicolumn{4}{c}{\textbf{Flickr30K}} & \multicolumn{4}{c}{\textbf{MSCOCO}} & \multicolumn{4}{c}{\textbf{FashionIQ}} \\
\cmidrule(lr){2-5} \cmidrule(lr){6-9} \cmidrule(lr){10-13}
\textbf{Method} & R@1 & R@5 & R@10 & MRR@10 & R@1 & R@5 & R@10 & MRR@10 & cR@1 & cR@5 & cR@10 & cMRR@10 \\
\midrule
\rowcolor{gray!8}
Dense & 77.10 & 93.70 & 97.30 & 84.04 & 46.30 & 76.90 & 87.40 & 58.94 & 28.57 & 59.65 & 71.59 & 41.94 \\
\midrule
\rowcolor{gray!10}
\multicolumn{13}{l}{\textit{Retention $R = 20\%$}} \\
FastV        & 61.80 & 89.00 & 94.50 & 72.99 & 35.90 & 70.10 & 83.70 & 49.84 & 11.48 & 34.20 & 48.97 & 21.84 \\
PyramidDrop  & 70.70 & 92.00 & 96.00 & 79.43 & 42.70 & \underline{73.70} & 85.00 & 55.41 & \underline{22.93} & 52.75 & 66.61 & 36.12 \\
SparseVLM    & 64.00 & 87.70 & 95.30 & 73.90 & 41.40 & 72.20 & \underline{86.70} & 54.64 & 21.83 & 53.58 & \underline{67.45} & 35.51 \\
DART         & 53.50 & 82.30 & 90.60 & 65.56 & 32.80 & 59.30 & 77.60 & 45.08 & 15.00 & 47.28 & 61.97 & 28.97 \\
VisionZip    & 63.50 & 89.10 & 94.30 & 74.03 & 38.10 & 68.20 & 81.10 & 51.07 & 17.01 & 45.03 & 59.44 & 29.41 \\
LowRes       & \underline{71.50} & \underline{92.20} & \underline{97.10} & \underline{80.03} & \underline{42.80} & 72.80 & 85.60 & \underline{55.64} & 22.75 & \underline{54.48} & 67.00 & \underline{36.55} \\
\textbf{RaDiCal} & \textbf{77.10} & \textbf{93.70} & \textbf{97.30} & \textbf{83.98} & \textbf{46.00} & \textbf{75.80} & \textbf{87.60} & \textbf{59.09} & \textbf{26.21} & \textbf{58.50} & \textbf{71.13} & \textbf{39.79} \\
\midrule
\rowcolor{gray!10}
\multicolumn{13}{l}{\textit{Retention $R = 10\%$}} \\
FastV        & 39.70 & 75.10 & 85.40 & 54.40 & 25.50 & 55.50 & 74.30 & 38.78 & 6.88 & 28.62 & 45.53 & 16.66 \\
PyramidDrop  & \underline{70.40} & \underline{92.40} & \underline{96.20} & \underline{79.28} & \underline{42.70} & \underline{73.50} & \underline{85.20} & \underline{55.33} & 19.45 & \underline{49.03} & \underline{63.25} & \underline{32.44} \\
SparseVLM    & 51.90 & 80.30 & 89.30 & 64.03 & 34.80 & 65.30 & 80.80 & 47.71 & \underline{19.70} & 48.55 & 61.68 & 32.22 \\
DART         & 39.30 & 70.60 & 81.90 & 52.44 & 21.80 & 50.90 & 70.10 & 34.43 & 13.60 & 41.96 & 57.56 & 25.98 \\
VisionZip    & 52.60 & 82.50 & 90.40 & 65.06 & 30.20 & 61.30 & 77.30 & 43.53 & 11.35 & 34.85 & 50.28 & 22.11 \\
LowRes       & 63.90 & 88.50 & 95.70 & 74.50 & 39.10 & 69.50 & 83.50 & 52.01 & 18.78 & 48.96 & 62.62 & 31.92 \\
\textbf{RaDiCal} & \textbf{76.00} & \textbf{92.80} & \textbf{96.90} & \textbf{82.85} & \textbf{45.10} & \textbf{76.10} & \textbf{88.50} & \textbf{58.00} & \textbf{24.78} & \textbf{56.84} & \textbf{69.84} & \textbf{37.99} \\
\bottomrule
\end{tabular}
\caption{Main reranking results across three benchmarks at two retention budgets. FashionIQ reports conditional metrics (\S\ref{sec:setup}). \textbf{Bold}: best pruning method; \underline{underline}: second-best. All values in \%.}
\label{tab:main_results}
\end{table*}

\paragraph{Uniform budget realization.}
\label{sec:uniform_budget}

Given selected layers $\mathcal{L}^* = (l_1, \dots, l_K)$ and global keep ratio $R$, we use the same per-layer keep ratio at every selected layer:
\begin{equation}
\text{keep}(l_h) = R^{1/K}, \qquad h=1,\dots,K.
\end{equation}
Sequential application yields realized retention $\prod_{h=1}^{K}\text{keep}(l_h)=R$; integer rounding applies conservatively at each layer.

\subsection{Unified Pipeline}
\label{sec:pipeline}

The framework separates into a one-time offline pass (per model and budget) and a per-inference online pass (per query $\times$ candidate set). The complete procedure is given in Figure~\ref{fig:pipeline} and Algorithm~\ref{alg:trust_pruning} (Appendix~\ref{app:algorithm}), which makes the token state explicit across pruning layers.

The offline schedule ($\mathcal{L}^*$, keep) is fixed for a given model and budget, matching the deployment profile of prior layer-selection methods~\citep{fastv,coast}. The external keep ratio $R$ controls the evaluation budget, while $K$ and the layer-gap guard are fixed deployment choices and are not tuned per dataset.

\section{Experimental Results}
\label{sec:experiments}

\subsection{Experimental Setup}
\label{sec:setup}

We evaluate RaDiCal on Qwen3-VL-4B-Instruct~\cite{qwen3vl} as the listwise reranker, with candidates retrieved by Qwen3-VL-Embedding-2B~\cite{qwen3vl_embedding}. Model generalization---to the 8B variant and to a different VLM architecture (InternVL2.5-8B~\cite{internvl25})---is examined in \S\ref{sec:cross_model}.

We use three benchmarks. Flickr30K~\cite{flickr30k} and MSCOCO~\cite{mscoco} follow the Karpathy test split~\cite{karpathy_split} (1{,}000 queries $\times$ 20 candidates each). FashionIQ~\cite{fashioniq} is a composed image retrieval benchmark where each query combines a reference image with a textual modification (validation split; 6{,}016 queries $\times$ 20 candidates drawn from a top-50 retrieval pool). For FashionIQ, we report conditional metrics computed over the 1{,}599 queries whose ground-truth target appears in the first-stage retriever's top-20 candidate set, isolating reranking quality from first-stage retrieval coverage. Appendix~\ref{app:impl_details} details how DTI adapts to FashionIQ's composed-query format.

We compare against six visual token reduction baselines---FastV~\cite{fastv}, PyramidDrop~\cite{pyramiddrop}, SparseVLM~\cite{sparsevlm}, DART~\cite{dart}, VisionZip~\cite{visionzip}, and LowRes (which simply reduces input image resolution)---all re-implemented within the same Qwen3-VL framework to ensure a fair comparison. Dense (no pruning) serves as the unpruned reference. Appendix~\ref{app:impl_details} details baseline descriptions and hyper-parameter configurations.

We report MRR@10 and R@1/5/10 on Flickr30K and MSCOCO, and their conditional counterparts (cR@$k$, cMRR@10) on FashionIQ; Appendix~\ref{app:fashioniq_unconditional} additionally reports unconditional metrics over all queries. We evaluate at two retention budgets: $R{=}20\%$ (primary) and $R{=}10\%$ (stress test); the global budget is realized by an equal per-layer keep ratio $R^{1/K}$ at each of the $K{=}4$ selected pruning layers (\S\ref{sec:scheduling}).

RaDiCal prunes at layers $[7, 22, 24, 29]$, selected by $\alpha$-Maximin (\S\ref{sec:maximin}), using the AttentionInfo saliency channel.

\subsection{Main Results}
\label{sec:main_results}

We evaluate whether RaDiCal preserves reranking quality under aggressive token pruning (Table~\ref{tab:main_results}).

At $R{=}20\%$, RaDiCal matches Dense quality---exceeding Dense MRR@10 on MSCOCO and matching it on Flickr30K (within 0.1\,pp). It ranks first in all 12 metric columns across all three benchmarks, with a clear margin on MRR@10.

Halving the budget to $R{=}10\%$ separates structurally robust methods from fragile ones. RaDiCal ranks first in all 12 columns, with MRR@10 within 1.2\,pp of Dense on Flickr30K and MSCOCO, and still above every baseline's $R{=}20\%$ score on all three benchmarks, while single-layer baselines (FastV, DART) degrade by an order of magnitude more. This stability shows that RaDiCal's token selection captures ranking-relevant structure rather than exploiting visual redundancy.

On FashionIQ, RaDiCal again achieves the highest cMRR@10 among pruning methods at both retention budgets. The method ranking is consistent across all three benchmarks: RaDiCal leads while single-layer baselines degrade most, extending to composed retrieval.

\begin{figure}[!t]
\centering
\includegraphics[width=\columnwidth]{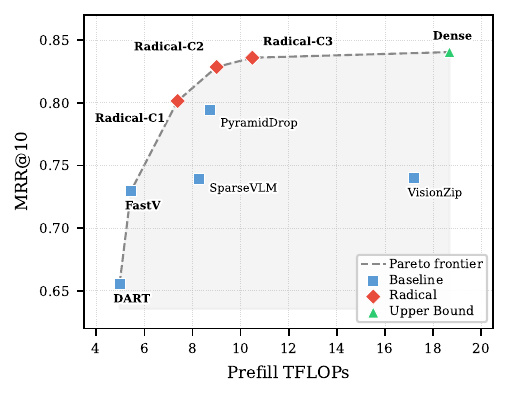}
\caption{Quality--efficiency Pareto frontier (Flickr30K, $R{=}20\%$).}
\label{fig:pareto}
\end{figure}

\begin{figure*}[!t]
\centering
\includegraphics[width=\textwidth]{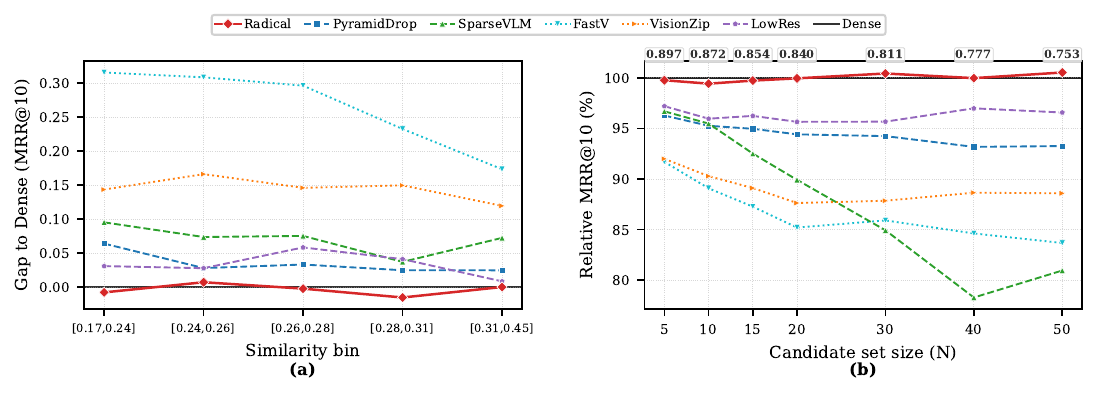}
\caption{Analysis: (a) performance gap to Dense across five candidate-similarity quintiles; (b) relative MRR@10 (Dense${}=100\%$) as candidate set size $N$ varies from 5 to 50.}
\label{fig:analysis}
\end{figure*}

\subsection{Efficiency Analysis}
\label{sec:efficiency}

\begin{table}[!t]
\centering
\small
\setlength{\tabcolsep}{3pt}
\renewcommand{\arraystretch}{1.05}
\resizebox{\columnwidth}{!}{%
\begin{tabular}{@{}ll rrr@{}}
\toprule
\textbf{Model} & \textbf{Method}
  & \textbf{Rel$_\text{D}$(\%)} & \textbf{FLOPs$\downarrow$} & \textbf{Speedup} \\
\midrule
\rowcolor{gray!10}
  & Dense       & 100.0         & --             & 1.00$\times$ \\
  & RaDiCal     & \textbf{99.9} & 43.8\%         & \textbf{1.28$\times$} \\
  & PyramidDrop & 94.5          & 53.3\%         & 0.76$\times$ \\
  & SparseVLM   & 87.9          & 55.7\%         & 1.12$\times$ \\
\multirow{-5}{*}{Qwen3-VL-4B}
  & FastV       & 86.9          & \textbf{70.8\%} & 1.09$\times$ \\
\midrule
\rowcolor{gray!10}
  & Dense       & 100.0         & --             & 1.00$\times$ \\
  & RaDiCal     & \textbf{99.3} & 39.2\%         & 1.45$\times$ \\
  & PyramidDrop & 95.1          & 55.6\%         & 1.68$\times$ \\
  & SparseVLM   & 94.3          & 60.7\%         & 1.46$\times$ \\
\multirow{-5}{*}{InternVL2.5-8B}
  & FastV       & 86.1          & \textbf{71.1\%} & \textbf{2.00$\times$} \\
\bottomrule
\end{tabular}}
\caption{Quality and measured efficiency across two VLM architectures (Flickr30K, $R{=}20\%$). Rel$_\text{D}$: MRR@10 relative to Dense. FLOPs$\downarrow$ is analytical; speedup is measured end-to-end. Rows are ordered by Rel$_\text{D}$; the best pruning result per model and metric is \textbf{bolded}. Full metrics in Appendix~\ref{app:efficiency}.}
\label{tab:eff_summary}
\end{table}

The quality advantage translates into a favorable quality--efficiency trade-off: RaDiCal lies on the Pareto frontier and dominates baselines in MRR@10 at comparable compute (Figure~\ref{fig:pareto}).
At $R{=}20\%$ on Qwen3-VL-4B, RaDiCal saves over 40\% of Dense FLOPs; reducing to $R{=}10\%$ pushes savings past 50\% with minimal additional degradation (per-method TFLOPs in Table~\ref{tab:eff_full}).
RaDiCal variants span a wide band of the frontier, giving practitioners a continuous quality--efficiency operating range rather than a single operating point.
RaDiCal is the only method above 99\% of Dense MRR@10 on both backbones, delivering 1.28--1.45$\times$ measured end-to-end speedups (Table~\ref{tab:eff_summary}). Analytical compression does not predict wall-clock gain: PyramidDrop saves more FLOPs than RaDiCal on both backbones yet achieves only 0.76$\times$ speedup on Qwen3-VL-4B---slower than Dense---while FastV, despite removing the most FLOPs on both backbones, yields the lowest ranking quality. At the smallest analytical compression of any pruning method, rank-aware selection, not aggressive token removal, best preserves ranking under a latency budget. Appendix~\ref{app:efficiency} reports full per-dataset and wall-clock metrics.

\subsection{Ablation Study}
\label{sec:ablation}

\begin{table}[!t]
\centering
\small
\begin{tabular}{@{}l cc@{}}
\toprule
\textbf{Variant} & \textbf{Flickr30K} & \textbf{MSCOCO} \\
\midrule
Full RaDiCal & 83.98 & 59.09 \\
w/o DTI ($\alpha{=}0$) & 82.98\,{\scriptsize\textcolor{red!70!black}{($-$1.00)}} & 58.39\,{\scriptsize\textcolor{red!70!black}{($-$0.70)}} \\
Fixed fusion ($\alpha{=}0.5$) & 82.65\,{\scriptsize\textcolor{red!70!black}{($-$1.33)}} & 58.15\,{\scriptsize\textcolor{red!70!black}{($-$0.94)}} \\
w/o $\alpha$-Maximin & 82.01\,{\scriptsize\textcolor{red!70!black}{($-$1.97)}} & 57.70\,{\scriptsize\textcolor{red!70!black}{($-$1.39)}} \\
w/o Geometric & 83.26\,{\scriptsize\textcolor{red!70!black}{($-$0.72)}} & 58.58\,{\scriptsize\textcolor{red!70!black}{($-$0.51)}} \\
DTI-only ($\alpha{=}1$) & 83.11\,{\scriptsize\textcolor{red!70!black}{($-$0.87)}} & 58.48\,{\scriptsize\textcolor{red!70!black}{($-$0.61)}} \\
QRel-only ($\alpha$-Maximin) & 83.24\,{\scriptsize\textcolor{red!70!black}{($-$0.74)}} & 58.57\,{\scriptsize\textcolor{red!70!black}{($-$0.52)}} \\
CCU-only ($\alpha$-Maximin) & 81.86\,{\scriptsize\textcolor{red!70!black}{($-$2.12)}} & 57.60\,{\scriptsize\textcolor{red!70!black}{($-$1.49)}} \\
w/o Token Scoring & 82.00\,{\scriptsize\textcolor{red!70!black}{($-$1.98)}} & 57.70\,{\scriptsize\textcolor{red!70!black}{($-$1.39)}} \\
\midrule
Dense & 84.04\,{\scriptsize\textcolor{green!50!black}{($+$0.06)}} & 58.94\,{\scriptsize\textcolor{red!70!black}{($-$0.15)}} \\
\bottomrule
\end{tabular}
\caption{Component removal ablation (MRR@10 (\%), $R{=}20\%$). Subscripts show the delta relative to Full RaDiCal (\textcolor{red!70!black}{red}: degradation; \textcolor{green!50!black}{green}: improvement).}
\label{tab:ablation}
\end{table}

We isolate each component's contribution by removing it and measuring the resulting degradation.
The component hierarchy is clear (Table~\ref{tab:ablation}): the three largest degradations on both datasets---CCU-only, w/o Token Scoring, and w/o $\alpha$-Maximin---cluster tightly, confirming that \emph{how} to score tokens and \emph{where} to prune carry roughly equal weight. Notably, fixed-weight mixing ($\alpha{=}0.5$) underperforms disabling DTI entirely ($\alpha{=}0$) on both datasets, indicating that the benefit of DTI depends on \emph{when} it is applied; adaptive trust calibration is necessary to realize the gain. Neither signal suffices alone: DTI-only and saliency-only both underperform the calibrated combination, confirming that the two channels are complementary rather than redundant.

Appendices~\ref{app:layer_selection} and~\ref{app:k_sensitivity} report design-choice justifications including alternative layer selection strategies and $K$ sensitivity. Appendix~\ref{app:case_study} visualizes these signal differences on representative queries.
Candidate-level resolution allocation---even guided by the same DTI signal---falls 3.70--4.43\,pp short of RaDiCal on Flickr30K at matched total budget: token-level spatial selection is essential (Appendix~\ref{app:dynres}).

\subsection{Analysis}
\label{sec:analysis}

\subsubsection{Candidate Similarity Impact}
\label{sec:similarity}

Listwise reranking becomes most challenging when candidates are visually similar, because shared visual patterns are more likely to inflate saliency for order-neutral tokens; we now test whether token pruning selectively degrades on such hard cases.
On raw MRR@10, RaDiCal follows the same decline pattern as Dense as candidate similarity increases; in the gap-to-Dense view (Figure~\ref{fig:analysis}\,a), the gap remains near zero across all five similarity bins, indicating that pruning preserves the information needed to distinguish similar candidates. In other words, visual ambiguity lowers absolute ranking quality without magnifying the cost of pruning.

Together, these results support the rank-discriminative criterion (Definition~\ref{def:rank_discriminative}): the discriminative signal compensates for saliency degradation in hard candidate sets. Appendix~\ref{app:case_study} illustrates this on a high-similarity example.

\subsubsection{Candidate Set Size Sensitivity}
\label{sec:candidate_size}

Candidate list length varies widely across retrieval pipelines, making robustness to $N$ a practical requirement.
As candidate set size $N$ varies from 5 to 50, RaDiCal's relative MRR@10 remains virtually flat---spanning barely 1\,pp around Dense even as Dense quality itself drops substantially---while baselines such as SparseVLM and FastV degrade by an order of magnitude more (Figure~\ref{fig:analysis}\,b).
Because DTI's global prototype adapts naturally to the candidate pool size, the pruning criterion remains well-calibrated across deployment scales, eliminating the need for per-$N$ retuning.

\subsection{Model Generalization}
\label{sec:cross_model}

\begin{table}[!t]
\centering
\small
\setlength{\tabcolsep}{3pt}
\renewcommand{\arraystretch}{1.05}
\begin{tabular}{@{}ll ccc@{}}
\toprule
\textbf{Model} & \textbf{Method}
  & \textbf{MRR@10} & \textbf{R@1} & \textbf{R@10} \\
\midrule
\rowcolor{gray!10}
  & Dense       & 86.93 & 80.30 & 98.10 \\
  & RaDiCal     & \textbf{87.14} & \textbf{80.40} & \textbf{98.30} \\
  & PyramidDrop & 84.39 & 77.00 & 97.00 \\
  & SparseVLM   & 85.51 & 75.80 & 97.50 \\
\multirow{-5}{*}{Qwen3-VL-8B}
  & FastV       & 61.76 & 50.30 & 87.60 \\
\midrule
\rowcolor{gray!10}
  & Dense       & 32.05 & 14.50 & 72.40 \\
  & RaDiCal     & \textbf{31.81} & \textbf{13.40} & 73.80 \\
  & PyramidDrop & 30.49 & 12.20 & \textbf{74.10} \\
  & SparseVLM   & 30.23 & 12.50 & 71.90 \\
\multirow{-5}{*}{InternVL2.5-8B}
  & FastV       & 27.59 & 10.10 & 69.30 \\
\bottomrule
\end{tabular}
\caption{Model generalization (Flickr30K, $R{=}20\%$).
Qwen3-VL-8B uses layers $[4,7,21,29]$;
InternVL2.5-8B uses layers $[8,22,25,27]$,
both auto-selected by $\alpha$-Maximin.
Best pruning method per group is \textbf{bolded}. All values in \%.}
\label{tab:generalization}
\end{table}

We now test whether the advantage transfers to a larger model (Qwen3-VL-8B) and a different VLM architecture (InternVL2.5-8B).
RaDiCal exceeds Dense MRR@10 on the larger Qwen3-VL-8B (Table~\ref{tab:generalization}), and retains over 99\% of Dense quality on InternVL2.5-8B.
$\alpha$-Maximin automatically selects a distinct layer schedule for each model (see caption of Table~\ref{tab:generalization}) without manual tuning.
The method ranking is stable across both architectures: RaDiCal leads on MRR@10 while FastV degrades most.
The advantage also survives a change of first-stage retriever: RaDiCal matches or exceeds Dense across three substantially different candidate pools (Qwen, Jina, SigLIP2) over five seeds, with pairwise pool overlap below 40\% (Appendix~\ref{app:retriever_robustness}).

\section{Conclusion}
\label{sec:conclusion}

We showed that attention saliency is systematically misaligned with ranking contribution in VLM-based listwise reranking, and that normalized attention entropy ($H_{\text{norm}}$) diagnoses where saliency becomes trustworthy. RaDiCal exploits this insight to unify token scoring, trust calibration, and layer scheduling under a single entropy-derived coordinate---matching Dense on Flickr30K and exceeding it on MSCOCO at a 20\% token budget with measured speedups of up to 1.45$\times$, and transferring across datasets, model scales, and architectures without retuning. The ablation confirms that \emph{where} to prune matters as much as \emph{how} to score. Integrating DTI-style discriminative priors into the training loop is a promising direction for further improving calibration fidelity across diverse retrieval scenarios.

\section*{Limitations}
\label{sec:limitations}
Our work has several limitations.
First, RaDiCal is designed for listwise reranking, where multiple candidates provide the cross-candidate contrast that DTI exploits; its applicability to other multi-image tasks (e.g., visual comparison, multi-image VQA) has not been evaluated.
Second, all components are training-free, and we do not investigate whether learning the calibration weight $\alpha$ or the DTI scoring weights in a task-specific manner could yield further gains.
Third, the method assumes softmax-based attention, so it may not generalize to architectures with alternative attention mechanisms such as linear attention or state-space models.
Finally, although our evaluation spans three reranking benchmarks, six single-image benchmarks (Appendix~\ref{app:vqa_transfer}), three first-stage retrievers, and two VLM families, we have not validated on a wider range of retrieval domains or modalities.

\section*{Ethical Considerations}
This work collects no new data or human-subject annotations; experiments use public benchmarks and pretrained models subject to their original licenses. RaDiCal reduces computation, but we do not measure energy use or carbon emissions, nor mitigate inherited biases, privacy risks, or hallucinations. Because pruning may alter candidate exposure and affect underrepresented groups, subgroup evaluation and human oversight are recommended before high-stakes deployment.

\paragraph{Use of AI Assistants.}
We used a large language model (LLM) in our synthesis pipeline for caption editing and quality filtering. During manuscript preparation, we also used AI assistants for language editing and proofreading. All research ideas, experimental designs, analyses, interpretations, and claims are the authors' own; the authors reviewed and take responsibility for the final manuscript.

\section*{Acknowledgments}
This work was sponsored by the CCF-Tencent Rhino-Bird Open Research Fund (No.~CCF-Tencent RAGR20250119), and was also supported by the Guangdong Basic and Applied Basic Research Foundation (No.~2025A1515010304), the Guangdong Province Project (No.~2024QN11X088), and the Guangzhou Science and Technology Planning Project (No.~2025A03J4491).

\bibliography{references}

\clearpage
\appendix

\section{Code Availability}
\label{app:code}

The source code for RaDiCal, including the entropy-calibrated schedule construction and DTI scoring, is publicly available at \url{https://github.com/ssui-liu/RaDiCal}.

\begin{table*}[t]
\centering
\small
\begin{tabular}{@{}l l r r l l@{}}
\toprule
\textbf{Dataset} & \textbf{Split} & \textbf{Queries} & \textbf{Cand/Q} & \textbf{Task} & \textbf{Metrics} \\
\midrule
Flickr30K~\citep{flickr30k} & Karpathy test & 1{,}000 & 20 & Image-text reranking & R@$k$, MRR@10 \\
MSCOCO~\citep{mscoco}       & Karpathy test & 1{,}000 & 20 & Image-text reranking & R@$k$, MRR@10 \\
FashionIQ~\citep{fashioniq} & Validation    & 6{,}016 & 20 & Composed reranking   & cR@$k$, cMRR@10 \\
\bottomrule
\end{tabular}
\caption{Dataset statistics for reranking evaluation.}
\label{tab:dataset_stats}
\end{table*}

\begin{table*}[t]
\centering
\small
\setlength{\tabcolsep}{4.7pt}
\begin{tabular}{@{}l l l l@{}}
\toprule
\textbf{Method} & \textbf{Pruning Layers} & \textbf{Keep Schedule} & \textbf{Notes} \\
\midrule
FastV~\citep{fastv}             & $[2]$          & Single-layer, keep $R$          & Early-layer attn.\ pruning \\
PyramidDrop~\citep{pyramiddrop} & $[7,15,23]$    & Cumul.\ $[.5,.25,R]$           & Progressive dropping \\
SparseVLM~\citep{sparsevlm}    & $[3,8,18]$     & Per-method schedule             & Text-cond.\ saliency \\
DART~\citep{dart}               & $[1]$          & Single-layer, keep $R$          & Duplication-aware (pivot dissimilarity) \\
VisionZip~\citep{visionzip}     & Pre-LLM        & \texttt{keep\_ratio}$\,{=}\,R$ & Encoder-side merging; FLOPs$\downarrow$ only 4--8\% \\
LowRes                          & N/A            & \texttt{max\_pixels} $\to R$    & Resolution reduction; TFLOPs not comparable \\
Dense                           & None           & 100\%                           & Unpruned reference \\
\bottomrule
\end{tabular}
\caption{Baseline configurations on Qwen3-VL-4B. All methods are applied without task-specific training.}
\label{tab:baseline_config}
\end{table*}

\section{Implementation Details}
\label{app:impl_details}

\paragraph{Datasets.}
Table~\ref{tab:dataset_stats} summarizes the evaluation benchmarks. Candidates are retrieved by Qwen3-VL-Embedding-2B~\citep{qwen3vl_embedding}: Flickr30K and MSCOCO use the top-20 from the full test index; FashionIQ draws the top-20 from a top-50 retrieval pool.

\paragraph{Baseline configurations.}
Table~\ref{tab:baseline_config} lists all baseline configurations on Qwen3-VL-4B. All methods are training-free, evaluated on the same model with the same global retention budget~$R$.

\paragraph{Efficiency measurement.}
All efficiency metrics are measured over 1{,}000 queries in batch-sequential mode on a single GPU, extracted from per-run telemetry. TFLOPs (theoretical FLOPs derived from token counts and model architecture) is our primary efficiency metric because it is hardware-agnostic and consistent with prior work~\citep{fastv,pyramiddrop,sparsevlm}. FLOPs Saved~\% is relative to the Dense baseline. KV Cache reports peak memory; latency is an end-to-end measurement including all online overhead.

\paragraph{FashionIQ conditional metrics.}
FashionIQ's first-stage retrieval does not guarantee the ground-truth target in the top-20 candidate set: only 1{,}599 of 6{,}016 queries have the target present. Conditional metrics (cR@$k$, cMRR@10) are computed exclusively over these 1{,}599 queries, isolating reranking quality from retrieval coverage. Flickr30K and MSCOCO do not require this treatment.

\paragraph{FashionIQ DTI adaptation.}
In composed image retrieval, each query pairs a reference image with a textual modification describing the desired change. During listwise reranking, the reference image enters the VLM prompt as protected query context: its visual tokens are excluded from pruning and from the DTI candidate token pool. Token pruning operates exclusively on candidate visual tokens. For DTI scoring, $q_{\mathrm{emb}}$ is derived from the textual modification, so that cross-candidate uniqueness (CCU) and query relevance (QRel) reflect how well each candidate token matches the requested attribute change rather than the reference appearance.

\paragraph{Reproducibility.}
Every experiment is repeated five times (seeds 42, 123, 456, 789, 1024); we report the mean across runs. Generation uses \texttt{max\_new\_tokens}$\,{=}\,$512. The framework is PyTorch with Transformers; eager attention is used at the four selected pruning layers to materialize the attention distribution required by AttentionInfo.

\paragraph{DTI computation.}
Let $V = \sum_{k=1}^{N} M_k$ be the total number of visual tokens in the candidate set. The global prototype $\bar{v}$ approximates cross-candidate uniqueness in $O(Vd)$, avoiding the $O(V^2d)$ cost of exact pairwise token comparison. An exact max-cosine variant is available as a transparent upper bound when candidate distributions are multimodal. DTI scores are min-max normalized within each query--candidate set before fusion.

\section{Diagnostic Protocol}
\label{app:diagnostic_protocol}

The permutation-corrected $z$-score $z_l$ used in \S\ref{sec:saliency_failure} is constructed as follows. For each query $q$ and layer $l$, let $S^{(l)}$ denote the saliency scores (attention weights) assigned to the visual tokens of all candidates. We compute a group-level ranking-contribution estimate $\text{RC}_g$ by masking each candidate's tokens in turn and measuring the resulting change in ranking score---this quantifies each token group's contribution to the output ranking.

The observed Spearman correlation $\rho_{\text{obs}}$ between $S^{(l)}$ and $\text{RC}_g$ is then compared against a null distribution obtained by randomly permuting the candidate group assignments. The $z$-score is:
\begin{equation}
z_l = \frac{\rho_{\text{obs}} - \mu_{\text{null}}}{\sigma_{\text{null}}},
\end{equation}
where $\mu_{\text{null}}$ and $\sigma_{\text{null}}$ are the mean and standard deviation of the null distribution, averaged across all queries.

The three attention-free reference baselines in Figure~\ref{fig:preliminary_curve} are computed purely from vision-encoder embeddings:
\begin{itemize}[leftmargin=*]
\item \textbf{U (Uniqueness):} Cross-candidate uniqueness score $1 - \cos(t_i^k, \bar{v})$, measuring each token's deviation from the global visual prototype.
\item \textbf{Q (Query Relevance):} $\max(\cos(t_i^k, q_{\text{emb}}), 0)$, measuring semantic relevance to the query.
\item \textbf{U$\times$Q:} The product of U and Q, combining both factors---this corresponds to the DTI score defined in \S\ref{sec:dti}.
\end{itemize}

\section{Complete Algorithm}
\label{app:algorithm}

Algorithm~\ref{alg:trust_pruning} details the full RaDiCal pipeline, separating the one-time offline schedule construction from the per-inference online token selection.

\begin{algorithm*}[t]
\caption{Trust-calibrated layer schedule and online token selection.}
\label{alg:trust_pruning}
\begin{algorithmic}[1]
\Require Model, candidate layer set $\mathcal{L}$, global keep ratio $R$, schedule size $K$, layer gap $g$, initial token set $T_0=\{1,\ldots,V\}$, ViT token embeddings, query embedding $q_{\text{emb}}$
\Ensure Ordered pruning layers $\mathcal{L}^*=(l_1,\ldots,l_K)$, per-layer keep ratios $\text{keep}(l_h)$ for $h=1,\ldots,K$, final retained token set $T_K$
\Statex
\Statex \textit{--- Offline schedule construction (once per model and budget) ---}
\For{each $l \in \mathcal{L}$}
    \State Compute $H_{\text{norm}}(l)$ and set $\alpha(l) \leftarrow (H_{\text{norm}}(l) - H_{\min}) / (1 - H_{\min})$
\EndFor
\State $\mathcal{L}^* \leftarrow \bigl\{\operatorname{TieBreak}\bigl(\arg\min_{l \in \mathcal{L}} \alpha(l)\bigr)\bigr\}$ \Comment{Anchor at most reliable layer}
\While{$|\mathcal{L}^*| < K$}
    \State $\mathcal{F} \leftarrow \{l \in \mathcal{L} \setminus \mathcal{L}^* : \forall l' \in \mathcal{L}^*,\; |l - l'| \geq g\}$
    \If{$\mathcal{F} = \emptyset$}
        \State $\mathcal{F} \leftarrow \mathcal{L} \setminus \mathcal{L}^*$ \Comment{Relax gap constraint}
    \EndIf
    \State $l^* \leftarrow \operatorname{TieBreak}\bigl(\arg\max_{l \in \mathcal{F}} \min_{l' \in \mathcal{L}^*} |\alpha(l) - \alpha(l')|\bigr)$ \Comment{$\alpha$-Maximin}
    \State $\mathcal{L}^* \leftarrow \mathcal{L}^* \cup \{l^*\}$
\EndWhile
\State Sort $\mathcal{L}^*$ by network depth, yielding $\mathcal{L}^* = (l_1, \ldots, l_K)$
\State $\text{keep}(l_h) \leftarrow R^{1/K}$ for $h=1,\ldots,K$
\Statex
\Statex \textit{--- Online pruning (per query $\times$ candidate set) ---}
\State Compute DTI scores $d_i = \text{DTI}(t_i^k)$ from ViT tokens and $q_{\text{emb}}$
\State $\hat{d}_i \leftarrow \text{MinMaxNorm}(\{d_i\}_{i \in T_0})$ \Comment{Normalize DTI once}
\For{$h = 1, \ldots, K$}
    \State $l \leftarrow l_h$
    \State Continue pruned forward pass to layer $l$ using active tokens $T_{h-1}$
    \State Compute AttentionInfo saliency $s_i^{(l, T_{h-1})}$ from attention over $T_{h-1}$
    \State $\hat{s}_i^{(l)} \leftarrow \text{MinMaxNorm}(\{s_i^{(l, T_{h-1})}\}_{i \in T_{h-1}})$; if $\max_i s_i = 0$: $\hat{s}_i^{(l)} \leftarrow 1/|T_{h-1}|$
    \State $f_i^{(l)} \leftarrow \max(\hat{d}_i, \epsilon)^{\alpha(l)} \cdot \max(\hat{s}_i^{(l)}, \epsilon)^{1 - \alpha(l)}$ for $i \in T_{h-1}$ \Comment{Geometric fusion}
    \State $T_h \leftarrow \operatorname{TopK}\bigl(T_{h-1},\; f^{(l)},\; \lceil \text{keep}(l) \cdot |T_{h-1}| \rceil\bigr)$
\EndFor
\State \Return $\mathcal{L}^*$, $\text{keep}$, $T_K$
\end{algorithmic}
\end{algorithm*}

\section{AttentionInfo Derivation}
\label{app:attinfo_derivation}

We derive the token-level AttentionInfo score from the layer-level normalized entropy. Let $T$ denote the current active token set with $V_T = |T|$, and let $p^{(l,T)}$ be the text-averaged attention distribution over active visual tokens at layer $l$. The KL gap over $T$ satisfies:
\begin{equation}
1 - H_{\text{norm}}(l,T) = \frac{\mathrm{KL}(p^{(l,T)} \| \mathcal{U}_T)}{\log V_T},
\end{equation}
where $\mathcal{U}_T$ is the uniform distribution over $T$. We decompose this scalar into per-token signed contributions:
\begin{align}
\tilde{I}^{(l,T)}_i
  &= p^{(l,T)}_i \cdot \log\!\bigl(V_T \cdot p^{(l,T)}_i\bigr), \\
\sum_{i \in T} \tilde{I}^{(l,T)}_i
  &= \mathrm{KL}\!\bigl(p^{(l,T)} \,\|\, \mathcal{U}_T\bigr).
\end{align}
Token $i$ has $\tilde{I}^{(l,T)}_i > 0$ if and only if $p^{(l,T)}_i > 1/V_T$---its attention weight exceeds the active-set uniform baseline, contributing positively to the layer's information gain. Conversely, $\tilde{I}^{(l,T)}_i < 0$ when the token receives below-average attention. Taking the positive part isolates tokens that carry information above the uniform prior:
\begin{align}
\mathrm{AttInfo}^{(l,T)}_i
  &= \max\!\bigl(\tilde{I}^{(l,T)}_i,\; 0\bigr) \nonumber \\
  &= \max\!\bigl(
       p^{(l,T)}_i \cdot \log(V_T \cdot p^{(l,T)}_i),\; 0
     \bigr).
\end{align}
Normalizing along the active token dimension yields $\text{Sal}^{(l,T)}_i \in [0,1]$. When no token exceeds the uniform baseline (all-zero AttInfo), Sal falls back to uniform---a natural graceful degradation that avoids degenerate pruning decisions at high-entropy layers.

\section{Qualitative Case Study}
\label{app:case_study}

We complement the quantitative results with two case studies that visualize the token retention patterns produced by different scoring signals and pruning methods. Both examples use the Flickr30K test set at $R{=}20\%$.

\subsection{Mechanism Retention Visualization}
\label{app:case_mechanism}

\begin{figure*}[t]
\centering
\includegraphics[width=\textwidth]{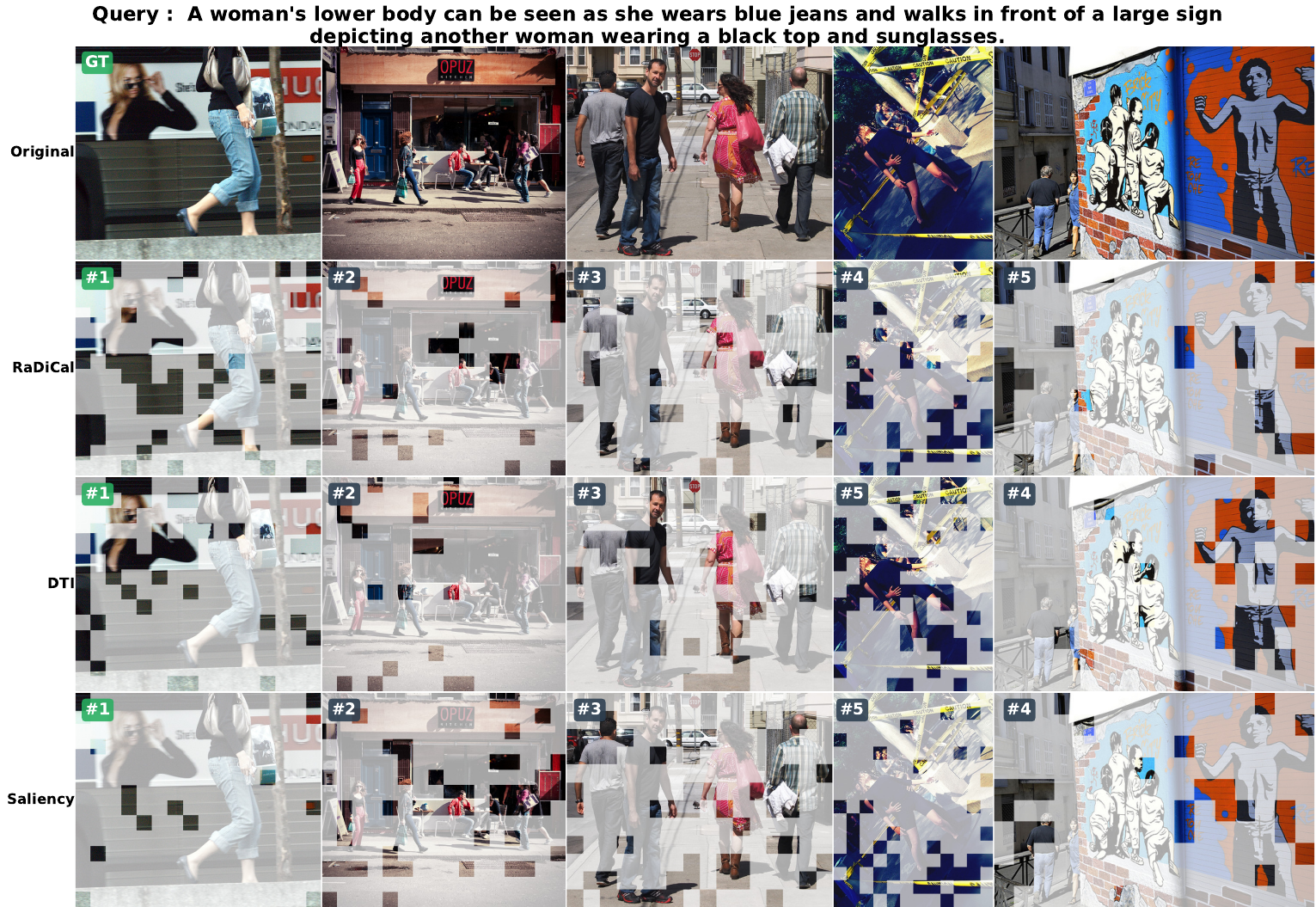}
\caption{Mechanism retention visualization on Flickr30K Query~270 (\emph{``A woman's lower body can be seen as she wears blue jeans and walks in front~\ldots''}). Each row applies a different scoring signal to the same five candidates (ground truth in column~1). Retained tokens appear at original brightness; pruned tokens are grayed out ($R{=}20\%$). Saliency-only retains visually prominent regions (signs, murals, high-contrast textures) regardless of query relevance. DTI-only focuses on query-relevant and cross-candidate distinctive evidence (blue jeans, lower body). RaDiCal combines both through entropy-calibrated fusion, preserving rank-discriminative cues with sufficient contextual evidence.}
\label{fig:case_mechanism}
\end{figure*}

Figure~\ref{fig:case_mechanism} compares three token-scoring signals---Saliency-only, DTI-only, and RaDiCal---on a query whose ranking hinges on identifying a woman's lower body wearing blue jeans.

\paragraph{Saliency bias.}
Saliency-only retains visually prominent regions such as storefronts, murals, and high-contrast background textures across all candidates. In candidates~4 and~5, strong colors and edges dominate the retention set despite being semantically irrelevant to the query. This directly illustrates the diagnostic in \S\ref{sec:saliency_failure}: attention saliency measures within-image prominence, not cross-candidate ranking evidence.

\paragraph{DTI focus.}
DTI-only shifts retention toward the ground-truth candidate's blue-jeans and lower-body regions---exactly the query-specified discriminative elements. On distractor candidates, DTI retains human-body tokens that could be query-relevant rather than entire background structures, consistent with Definition~\ref{def:rank_discriminative}: a token must be both query-relevant and cross-candidate distinctive to receive a high DTI score.

\paragraph{Calibrated fusion.}
RaDiCal inherits DTI's discriminative focus while AttentionInfo supplies layer-specific contextual structure, producing a more spatially coherent retention pattern than DTI-only without the background-attraction bias of Saliency-only. Notably, all three signals rank the ground truth first on this query; the value of the visualization is not a single-query outcome but the qualitative difference in \emph{what} each signal preserves---explaining why the calibrated combination yields stronger aggregate performance (Table~\ref{tab:ablation}).

\subsection{Cross-Method Retention Comparison}
\label{app:case_baseline}

\begin{figure*}[t]
\centering
\includegraphics[width=\textwidth]{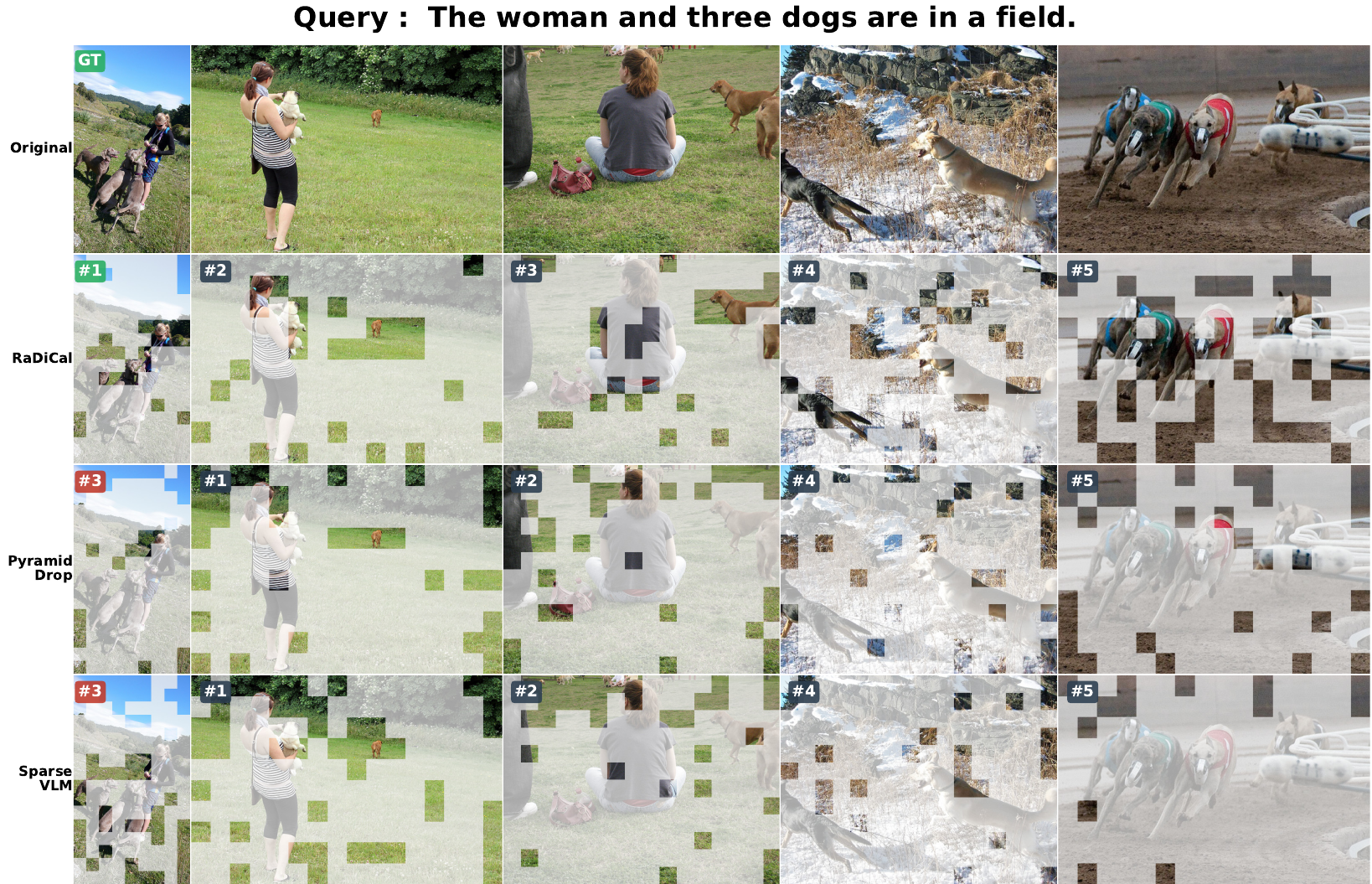}
\caption{Cross-method token retention on Flickr30K Query~658 (\emph{``The woman and three dogs are in a field.''}), a high-similarity case (sim\_bin${=}5$). Rows compare RaDiCal, PyramidDrop, and SparseVLM at $R{=}20\%$. RaDiCal correctly ranks the ground truth first, while PyramidDrop and SparseVLM both rank it third.}
\label{fig:case_baseline}
\end{figure*}

Figure~\ref{fig:case_baseline} compares RaDiCal against PyramidDrop and SparseVLM on a high-similarity query where all candidates share the visual pattern ``person~+~dog~+~outdoor.''

\paragraph{Query difficulty.}
The query requires satisfying three simultaneous constraints---a woman, three dogs, and a field---while lying in the hardest similarity quintile (sim\_bin${=}5$). Shared foreground objects inflate saliency uniformly across candidates, leaving little cross-candidate discriminative signal for attention-based methods.

\paragraph{RaDiCal.}
RaDiCal precisely retains tokens covering the three dogs' heads and bodies together with the woman's head in the ground-truth candidate---the key evidence distinguishing it from distractors that contain only one or two dogs. For clearly mismatched candidates (e.g., a dog-racing scene), RaDiCal allocates almost no token budget, concentrating resources on near-miss candidates that require fine-grained distinction.

\paragraph{PyramidDrop.}
PyramidDrop's retention forms a near-uniform grid, spreading tokens across undifferentiated grassland and keeping an approximately equal budget (${\sim}$19--20\%) per candidate regardless of query relevance---unable to exploit cross-candidate differences.

\paragraph{SparseVLM.}
SparseVLM commits irreversible pruning at shallow layers (L3,~L8) where $H_{\text{norm}}{\approx}1$ and attention is nearly uniform. This removes critical woman-head tokens from the ground-truth candidate and produces a query-agnostic, position-biased retention pattern, consistent with the shallow-layer unreliability diagnosed in \S\ref{sec:saliency_failure}.

\section{Layer Selection Strategy Comparison}
\label{app:layer_selection}

\begin{table*}[t]
\centering
\small
\begin{tabular}{@{}l l l r@{}}
\toprule
\textbf{Strategy} & \textbf{Layers} & \textbf{$\alpha$ Vector} & \textbf{MRR@10} \\
\midrule
$\alpha$-Maximin (Ours) & [7, 22, 24, 29] & [0.84, 0.43, 0.22, 0.00] & \textbf{83.98} \\
Cum.\ Info Equipart.    & [12, 21, 26, 30] & [0.83, 0.50, 0.03, 0.20] & \underline{83.82} \\
Max Gradient            & [17, 24, 27, 33] & [0.50, 0.22, 0.04, 0.10] & 83.40 \\
Alpha Quantile          & [15, 21, 25, 34] & [0.67, 0.50, 0.18, 0.39] & 83.16 \\
Deep-only               & [22, 24, 29]      & [0.43, 0.22, 0.00]       & 82.60 \\
Random 4 layers         & random           & random                    & 82.00 \\
Local Extrema           & [7, 14, 26, 29]  & [0.84, 0.82, 0.03, 0.00] & 81.97 \\
\bottomrule
\end{tabular}
\caption{Layer selection strategy comparison (Flickr30K, $R{=}20\%$, SparseVLM saliency channel, $K{=}4$). All MRR@10 values in \%.}
\label{tab:layer_selection}
\end{table*}

The key advantage of $\alpha$-Maximin is explicit diversity maximization in $\alpha$-space (Table~\ref{tab:layer_selection}): the selected layers $[7, 22, 24, 29]$ yield $\alpha$ values $[0.84, 0.43, 0.22, 0.00]$, spanning the three phases of the $H_{\text{norm}}$ curve: DTI-dominant ($\alpha{=}0.84$, shallow plateau) through mixed ($\alpha{=}0.43$, descending slope) and saliency-leaning ($\alpha{=}0.22$, lower slope) to saliency-only ($\alpha{=}0.00$, deep valley). Each pruning stage thus applies a qualitatively different scoring criterion, ensuring that the four budget-allocation decisions are complementary rather than redundant. The runner-up, Cumulative Information Equipartition (83.82), confirms that information-theoretic guidance is valuable, but its $\alpha$ vector $[0.83, 0.50, 0.03, 0.20]$ clusters two values near zero, yielding less uniform coverage than Maximin.

Conversely, Local Extrema selects layers whose $\alpha$ values cluster at the endpoints: $[0.84, 0.82, 0.03, 0.00]$ places two layers in the DTI-dominant regime, producing redundant decisions and the lowest structured-strategy MRR@10 (81.97)---comparable to Random (82.00). Removing the sole shallow DTI-dominant layer (L7, $\alpha{=}0.84$) from the Maximin set reduces MRR@10 by 1.38\,pp (Deep-only: 82.60), confirming that the shallow high-$\alpha$ stage is essential for full $\alpha$-space coverage. This gap is consistent in direction with the $-$1.97\,pp ablation drop reported in Section~\ref{sec:ablation} when replacing $\alpha$-Maximin (as defined in Section~\ref{sec:method}) with a na\"ive layer assignment.

\section{K Sensitivity}
\label{app:k_sensitivity}

\begin{figure}[t]
\centering
\includegraphics[width=\columnwidth]{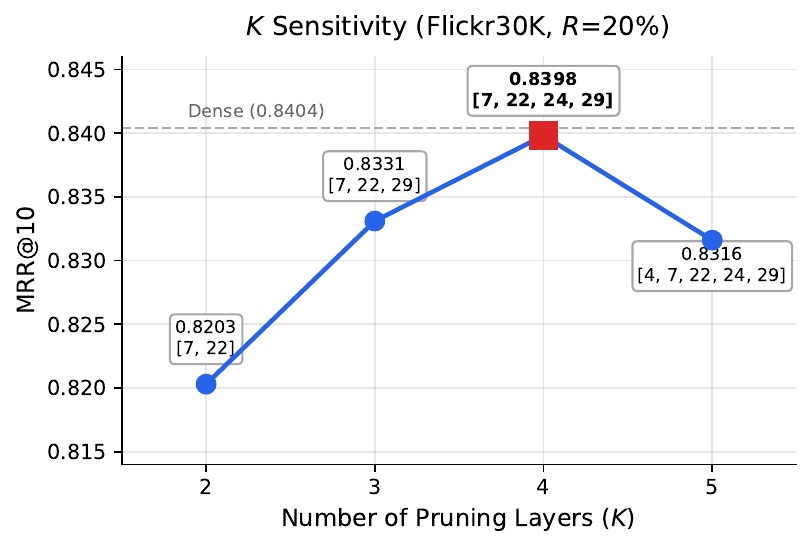}
\caption{$K$ sensitivity: MRR@10 as a function of the number of pruning stages (Flickr30K, $R{=}20\%$, AttentionInfo channel). Each point is annotated with the $\alpha$-Maximin--selected layers. The optimal $K{=}4$ (red square) nearly matches Dense (dashed line).}
\label{fig:k_sensitivity}
\end{figure}

The MRR@10 trajectory across $K{=}2,3,4,5$ forms a concave curve peaking at $K{=}4$ (Figure~\ref{fig:k_sensitivity}), confirming that four pruning stages is the empirical optimum rather than an arbitrary choice. Most of the gain is captured by $K{=}3$; $K{=}4$ supplies a further refinement that nearly closes the gap to Dense, while extending to $K{=}5$ introduces L4 ($\alpha{=}0.63$), where attention has not yet specialized sufficiently, and quality reverses. All layers are selected by $\alpha$-Maximin (\S\ref{sec:method}).

\paragraph{Layer-gap sensitivity.}
Table~\ref{tab:layer_gap} sweeps the minimum layer gap $g$ in $\alpha$-Maximin from 1 to 6 at $K{=}4$.

\begin{table}[h]
\centering
\footnotesize
\setlength{\tabcolsep}{2pt}
\begin{tabular}{@{}c l rr@{}}
\toprule
\textbf{$g$} & \textbf{Selected Layers} & \textbf{Flickr30K} & \textbf{MSCOCO} \\
\midrule
\textbf{1 or 2} (default) & \textbf{[7, 22, 24, 29]} & \textbf{83.98} & \textbf{59.09} \\
3 & [4, 7, 22, 29] & 82.56\,{\scriptsize($-$1.42)} & 57.70\,{\scriptsize($-$1.39)} \\
4--6 & [7, 15, 22, 29] & 81.82\,{\scriptsize($-$2.16)} & 56.79\,{\scriptsize($-$2.30)} \\
\bottomrule
\end{tabular}
\caption{Layer-gap sensitivity (Qwen3-VL-4B, $K{=}4$, $R{=}20\%$). MRR@10 (\%).}
\label{tab:layer_gap}
\end{table}

\noindent
$g \leq 2$ yields the highest mean MRR@10 on both datasets; over-spacing removes L24 from the schedule, breaking the dense coverage of the steep $H_{\mathrm{norm}}$ descent between L22 and L29, and degrading performance by up to 2.30\,pp.

\section{Efficiency}
\label{app:efficiency}

\paragraph{RaDiCal variants in Figure~\ref{fig:pareto}.}
The three RaDiCal configurations on the Pareto frontier are generated by varying the schedule size~$K$ passed to $\alpha$-Maximin (\S\ref{sec:scheduling}), which indirectly controls how deep the pruning schedule extends:

\begin{itemize}[nosep,leftmargin=1.2em]
  \item \textbf{RaDiCal-C1} ($K{=}3$, $R{=}20\%$): $\alpha$-Maximin selects layers $[7,22,29]$; per-layer keep ratio $R^{1/3}{\approx}0.585$, MRR@10${=}83.31$. %
  \item \textbf{RaDiCal-C2} ($K{=}4$, $R{=}10\%$): layers $[7,22,24,29]$; extending to layer~29 with a tighter budget gives 9.00~TFLOPs.
  \item \textbf{RaDiCal-C3} ($K{=}4$, $R{=}20\%$): same layers $[7,22,24,29]$; the standard budget yields 10.49~TFLOPs.
\end{itemize}

\noindent
A larger $K$ spreads the same global budget over more stages (per-layer keep $= R^{1/K}$ rises with $K$), so more tokens survive each layer; the deepest pruning layer is fixed by the anchor and does not move with $K$. With the layer schedule fixed, the global keep ratio~$R$ provides a second, orthogonal efficiency knob (C2 vs.\ C3).

\paragraph{Analytical metrics.}
Table~\ref{tab:eff_full} reports TFLOPs, FLOPs reduction, and KV-cache reduction for every method on both datasets at $R{=}20\%$ and $R{=}10\%$.

\begin{table*}[t]
\centering
\scriptsize
\setlength{\tabcolsep}{3pt}
\medskip
\textbf{(a) Flickr30K}
\vspace{2pt}

\begin{tabular}{@{}l rrr rrr@{}}
\toprule
& \multicolumn{3}{c}{\textbf{$R{=}20\%$}} & \multicolumn{3}{c}{\textbf{$R{=}10\%$}} \\
\cmidrule(lr){2-4} \cmidrule(lr){5-7}
\textbf{Method} & \textbf{TFLOPs} & \textbf{FLOPs$\downarrow$\%} & \textbf{KV$\downarrow$\%} & \textbf{TFLOPs} & \textbf{FLOPs$\downarrow$\%} & \textbf{KV$\downarrow$\%} \\
\midrule
Dense               & 18.68 & --     & --     & 18.68 & --     & --     \\
RaDiCal             & 10.49 & 43.8\% & 39.0\% & 9.00  & 51.8\% & 47.1\% \\
PyramidDrop         & 8.73  & 53.3\% & 50.1\% & 8.22  & 56.0\% & 53.1\% \\
SparseVLM           & 8.27  & 55.7\% & 51.9\% & 5.72  & 69.4\% & 66.4\% \\
FastV               & 5.45  & 70.8\% & 67.4\% & 4.04  & 78.3\% & 75.9\% \\
VisionZip           & 17.20 & 7.9\%  & 6.8\%  & 17.93 & 4.0\%  & 3.4\%  \\
DART                & 4.97  & 73.4\% & 70.0\% & 3.51  & 81.2\% & 78.7\% \\
\bottomrule
\end{tabular}

\medskip
\textbf{(b) MSCOCO}
\vspace{2pt}

\begin{tabular}{@{}l rrr rrr@{}}
\toprule
& \multicolumn{3}{c}{\textbf{$R{=}20\%$}} & \multicolumn{3}{c}{\textbf{$R{=}10\%$}} \\
\cmidrule(lr){2-4} \cmidrule(lr){5-7}
\textbf{Method} & \textbf{TFLOPs} & \textbf{FLOPs$\downarrow$\%} & \textbf{KV$\downarrow$\%} & \textbf{TFLOPs} & \textbf{FLOPs$\downarrow$\%} & \textbf{KV$\downarrow$\%} \\
\midrule
Dense               & 23.29 & --     & --     & 23.29 & --     & --     \\
RaDiCal             & 12.82 & 44.9\% & 39.5\% & 10.96 & 52.9\% & 47.7\% \\
PyramidDrop         & 10.67 & 54.2\% & 50.7\% & 10.03 & 56.9\% & 53.8\% \\
SparseVLM           & 3.81  & 83.6\% & 81.8\% & 3.33  & 85.7\% & 84.2\% \\
FastV               & 6.53  & 72.0\% & 68.3\% & 4.79  & 79.4\% & 76.8\% \\
VisionZip           & 21.39 & 8.1\%  & 6.8\%  & 22.35 & 4.1\%  & 3.4\%  \\
DART                & 5.93  & 74.5\% & 71.8\% & 4.13  & 82.3\% & 80.6\% \\
\bottomrule
\end{tabular}
\caption{Analytical efficiency on Qwen3-VL-4B at $R{=}20\%$ and $R{=}10\%$. LowRes excluded (resolution reduction, not token pruning).}
\label{tab:eff_full}
\end{table*}

\paragraph{Measurement protocol.}
We instrument per-query latency across all online components (DTI scoring, entropy-calibrated fusion, token selection) and compute speedup from total elapsed time, which also captures run-level overhead outside the timed regions. The $H_{\mathrm{norm}}$ profile and pruning schedule are computed offline once per model (\S\ref{sec:scheduling}) and amortized over all queries, as in prior layer-selection methods~\citep{fastv,coast}.

\paragraph{Scoring overhead.}
At matched budget on InternVL2.5-8B (39.2\% FLOPs$\downarrow$ for both), RaDiCal's per-query latency is no higher than score-free random pruning (3.81 vs.\ 3.89\,s/q), despite delivering meaningfully higher ranking quality (Table~\ref{tab:wallclock}). Online DTI and entropy scoring therefore add no measurable runtime cost.

\begin{table*}[t]
\centering
\scriptsize
\setlength{\tabcolsep}{4pt}
\begin{tabular}{@{}l rrrr rrrr@{}}
\toprule
& \multicolumn{4}{c}{\textbf{Qwen3-VL-4B}} & \multicolumn{4}{c}{\textbf{InternVL2.5-8B}} \\
\cmidrule(lr){2-5} \cmidrule(lr){6-9}
\textbf{Method}
& \textbf{Rel$_\text{D}$(\%)} & \textbf{FLOPs$\downarrow$} & \textbf{Lat.\ (s/q)} & \textbf{Speedup}
& \textbf{Rel$_\text{D}$(\%)} & \textbf{FLOPs$\downarrow$} & \textbf{Lat.\ (s/q)} & \textbf{Speedup} \\
\midrule
Dense        & 100.0 & --     & 4.477 & 1.00$\times$ & 100.0 & --     & 5.76 & 1.00$\times$ \\
RaDiCal      & 99.9  & 43.8\% & 3.472 & 1.28$\times$ & 99.3  & 39.2\% & 3.81 & 1.45$\times$ \\
Random       & 97.6  & 40.1\% & 4.180 & 1.07$\times$ & 96.1  & 39.2\% & 3.89 & 1.43$\times$ \\
PyramidDrop  & 94.5  & 53.3\% & 5.868 & 0.76$\times$ & 95.1  & 55.6\% & 3.25 & 1.68$\times$ \\
SparseVLM    & 87.9  & 55.7\% & 3.974 & 1.12$\times$ & 94.3  & 60.7\% & 3.80 & 1.46$\times$ \\
FastV        & 86.9  & 70.8\% & 4.085 & 1.09$\times$ & 86.1  & 71.1\% & 2.65 & 2.00$\times$ \\
\bottomrule
\end{tabular}
\caption{Measured efficiency, including the random-pruning control (Flickr30K, $R{=}20\%$, 1{,}000 queries $\times$ 20 candidates). Rel$_\text{D}$: MRR@10 relative to Dense. FLOPs$\downarrow$ is analytical; latency and speedup are measured end-to-end from total elapsed time.}
\label{tab:wallclock}
\end{table*}

\FloatBarrier
\section{Retriever Robustness}
\label{app:retriever_robustness}

We evaluate whether the advantage transfers across first-stage retrievers by testing three substantially different candidate pools---Qwen3-VL-Embedding-2B, Jina Embeddings v4, and SigLIP2-base-patch16-512 (pairwise Jaccard@20 $= 0.34$--$0.37$; RBO at $p{=}0.9 = 0.53$--$0.56$). We report Flickr30K ($R{=}20\%$) over five seeds; Rel$_\text{D}$ denotes MRR@10 relative to Dense (Table~\ref{tab:retriever_robustness}).

\begin{table}[h]
\centering
\small
\begin{tabular}{@{}l rrr@{}}
\toprule
\textbf{Method} & \textbf{Qwen} & \textbf{Jina} & \textbf{SigLIP2} \\
\midrule
Dense       & 84.04 & 84.90 & 85.07 \\
\textbf{RaDiCal} & \textbf{83.98} & \textbf{85.38} & \textbf{85.03} \\
PyramidDrop & 79.43 & 80.66 & 80.77 \\
SparseVLM   & 73.90 & 77.43 & 77.81 \\
\midrule
Rel$_\text{D}$ (\%) & 99.93 & 100.57 & 99.95 \\
\bottomrule
\end{tabular}
\caption{Retriever robustness (Flickr30K, $R{=}20\%$, 5 seeds). MRR@10 (\%).}
\label{tab:retriever_robustness}
\end{table}

RaDiCal matches or exceeds Dense across all three retrievers (99.93--100.57\% of Dense MRR@10), consistently leading PyramidDrop by 4.26--4.72\,pp. This follows from the design: DTI is pool-conditioned and recomputed per candidate set, while the pruning schedule stays model-specific and fixed.

\FloatBarrier
\section{Reranking-Aware Resolution Baselines}
\label{app:dynres}

To test a reranking-aware compression strategy at coarser granularity, we design two \emph{DynRes} baselines that allocate per-candidate image resolution (via \texttt{max\_pixels}) rather than performing token-level pruning. Both match RaDiCal's total visual-token budget at $R{=}20\%$.

\begin{itemize}[nosep,leftmargin=1.2em]
  \item \textbf{DynRes-QRel} allocates higher resolution to candidates with higher ViT-level query--candidate relevance.
  \item \textbf{DynRes-DTI} allocates by the full DTI signal (QRel $\times$ CCU), giving more resolution to candidates that are both query-relevant and cross-candidate distinctive.
\end{itemize}

\noindent
Each assigns a different \texttt{max\_pixels} per candidate via linear interpolation from a minimum to maximum resolution, calibrated so the total visual tokens match the RaDiCal budget. Neither baseline prunes tokens; all compression happens at the resolution level.

\begin{table}[h]
\centering
\small
\begin{tabular}{@{}l rr@{}}
\toprule
\textbf{Method} & \textbf{Flickr30K} & \textbf{MSCOCO} \\
\midrule
Dense               & 84.04 & 58.94 \\
\textbf{RaDiCal} ($R{=}20\%$) & \textbf{83.98} & \textbf{59.09} \\
DynRes-QRel         & 80.28 & 56.65 \\
LowRes (uniform)    & 80.03 & 55.64 \\
DynRes-DTI          & 79.55 & 55.94 \\
\bottomrule
\end{tabular}
\caption{DynRes resolution-allocation baselines vs.\ token-level pruning (MRR@10 (\%), $R{=}20\%$, Qwen3-VL-4B).}
\label{tab:dynres}
\end{table}

Token-level spatial selection proves essential (Table~\ref{tab:dynres}): RaDiCal leads the best DynRes variant by 2.4--3.7\,pp across both datasets. Both DynRes variants perform comparably to uniform low-resolution (LowRes), confirming that candidate-level resolution allocation---even guided by the same DTI signal---cannot substitute for within-image token selection.

\FloatBarrier
\section{FashionIQ: Conditional and Unconditional Metrics}
\label{app:fashioniq_unconditional}

The main results (Table~\ref{tab:main_results}) report conditional metrics computed over the 1{,}599 queries whose ground-truth target appears in the first-stage retriever's top-20 candidate set. Table~\ref{tab:fashioniq_both} additionally reports unconditional metrics computed over all 6{,}016 queries; method ordering is identical under both protocols.

\begin{table}[h]
\centering
\resizebox{\columnwidth}{!}{%
\begin{tabular}{@{}l rrrr@{}}
\toprule
\textbf{Method} & \textbf{cMRR@10} & \textbf{cR@10} & \textbf{uMRR@10} & \textbf{uR@10} \\
\midrule
Dense              & 41.94 & 71.59 & 11.17 & 19.13 \\
\textbf{RaDiCal} ($R{=}20\%$)  & \textbf{39.79} & \textbf{71.13} & \textbf{10.58} & \textbf{18.95} \\
LowRes             & 36.55 & 67.00 & 9.86  & 17.98 \\
PyramidDrop        & 36.12 & 66.61 & 9.62  & 17.69 \\
SparseVLM          & 35.51 & 67.45 & 9.47  & 18.04 \\
FastV              & 21.84 & 48.97 & 5.89  & 13.13 \\
\bottomrule
\end{tabular}}
\caption{FashionIQ conditional and unconditional metrics ($R{=}20\%$, Qwen3-VL-4B). All values in \%.}
\label{tab:fashioniq_both}
\end{table}

\FloatBarrier
\newpage
\section{Single-Image Transfer: VQA and Captioning}
\label{app:vqa_transfer}

CCU is inherently reranking-specific: without multiple candidates, cross-candidate uniqueness is undefined (Definition~\ref{def:rank_discriminative}). The remaining components---AttentionInfo saliency, $\alpha$ calibration, and $\alpha$-Maximin layer scheduling---are candidate-independent and can be applied to single-image tasks. To test whether these components transfer, we evaluate a $-$CCU variant (QRel $\times$ AttentionInfo with the same $\alpha$ calibration and layer scheduling) on six single-image benchmarks at $R{=}20\%$ on Qwen3-VL-4B (Table~\ref{tab:vqa_transfer}).

\begin{table}[h]
\centering
\footnotesize
\setlength{\tabcolsep}{2pt}
\begin{tabular}{@{}l rrrrr@{}}
\toprule
\textbf{Benchmark} & \textbf{Dense} & \textbf{Ours} & \textbf{PyrDrop} & \textbf{Sparse} & \textbf{FastV} \\
\midrule
GQA        & 54.29 & 53.88          & \textbf{53.90} & 48.06 & 46.68 \\
MME        & 80.37 & 80.29          & \textbf{80.79} & 77.93 & 76.16 \\
POPE-adv.  & 85.83 & \textbf{85.93} & 85.70          & 78.10 & 77.53 \\
COCO-Cap.  & 37.25 & \textbf{36.24} & 35.61          & 34.99 & 35.44 \\
DocVQA     & 90.92 & 78.25          & \textbf{86.35} & 55.53 & 49.09 \\
TextVQA    & 87.86 & 75.16          & \textbf{85.14} & 63.28 & 61.42 \\
\bottomrule
\end{tabular}
\caption{Single-image transfer ($-$CCU variant, $R{=}20\%$, Qwen3-VL-4B). Row metrics are exact match (GQA), overall accuracy (MME), accuracy (POPE-adv.), CIDEr$\times$100 (COCO-Cap.), ANLS (DocVQA), and VQA accuracy (TextVQA). Best pruning method per row is \textbf{bolded}.}
\label{tab:vqa_transfer}
\end{table}

\noindent
Even with its core cross-candidate term ablated, the $-$CCU variant leads on POPE-adv.\ and COCO-Caption, is within 0.02\,pp of PyramidDrop on GQA and 0.50\,pp on MME, and ranks above SparseVLM and FastV on all six tasks. PyramidDrop's margin is negligible on GQA and MME ($\le$0.5\,pp) but widens sharply on the two OCR-heavy benchmarks (DocVQA, TextVQA), where fine-grained text tokens dominate---a regime outside RaDiCal's listwise target.

\end{document}